\documentclass[a4paper,11pt]{article}
\usepackage[margin=2cm]{geometry}
\usepackage{epsfig}
\usepackage{hyperref}
\usepackage{authblk}

\title{Photoproduction of $J/\psi$ with neutron tagging in ultra-peripheral collisions of nuclei at RHIC and the LHC}

\author[1]{E. Kryshen}
\author[2]{M. Strikman}
\author[1]{M. Zhalov}
\affil[1]{Petersburg Nuclear Physics Institute, National Research Center ``Kurchatov Institute'', Gatchina, 188300, Russia}
\affil[2]{Department of Physics, The Pennsylvania State University, State College, PA 16802, USA}

\begin{document}
\maketitle

\begin{abstract}
We present predictions for the cross sections of the coherent and incoherent $J/\psi$ photoproduction in ultra-peripheral collisions at RHIC and at the LHC calculated for different classes of events depending on the presence of neutrons emitted by colliding nuclei. Since strong nucleus-nucleus interactions in UPCs are suppressed, it is usually assumed that neutrons at forward rapidities originate mainly from the electromagnetic dissociation of colliding nuclei caused by additional photon exchanges. This is a reasonable assumption for the coherent photoproduction where the state of the target nucleus remains intact. We consider additional sources of neutrons in the incoherent quasielastic and nucleon dissociative $J/\psi$ photoproduction and show that these processes significantly change probabilities of neutron emission compared to calculations when only neutrons from electromagnetic dissociation of nuclei are considered. Such studies should allow one to explore the dynamics of nuclear shadowing in the incoherent $J/\psi$ photoproduction down to $x \approx 10^{-5}$.

\end{abstract}

\section{Introduction}
During the last decade, ultra-peripheral collisions (UPCs) of heavy nuclei at the LHC have been extensively used for the coherent and incoherent $J/\psi$ photoproduction studies~\cite{Baltz:2007kq,Contreras:2015dqa}. Such measurements are also in progress at lower energies with the STAR detector at RHIC~\cite{STAR:2021wwq}. The main goal of these studies is to determine the energy and the momentum transfer dependencies of the coherent and incoherent $J/\psi$ photoproduction cross sections on nuclear targets which, within the perturbative QCD, are tightly related to the three-dimensional nuclear gluon density distributions. 

Coherent $J/\psi$ photoproduction on nuclear targets is studied in various phenomenological approaches based on elastic Glauber-like rescatterings of hadronic components of the photon~\cite{Klein:2016yzr}, Glauber-Gribov inelastic rescatterings within the leading twist approximation~\cite{Frankfurt:2011cs}, and in various implementations of the color-dipole approach~\cite{Cisek:2012yt,Goncalves:2014wna,Lappi:2013am,Cepila:2017nef,Luszczak:2019vdc,Bendova:2020hbb,Henkels:2020kju}. In particular, within the leading twist approximation (LTA) of nuclear gluon shadowing~\cite{Frankfurt:2011cs}, the coherent $J/\psi$ photoproduction cross section on nuclear target is proportional to the squared ratio of the nuclear $G_A(x,\mu^2)$ and the nucleon $G_N(x,\mu^2)$  densities of gluons carrying the fraction $x$ of the nucleon momentum at the hard scale $\mu$ characterizing the $J/\psi$ photoproduction process~\cite{Frankfurt:2003wv}.  Coherent $J/\psi$ measurements at RHIC and at the LHC should allow for the study of $G_A(x,\mu^2)$ behaviour in the range $10^{-5}<x<10^{-2}$ at $\mu^2\approx 3$\, GeV$^2$.

The study of incoherent $J/\psi$ photoproduction in UPCs also attracted significant theoretical interest in the past few years~\cite{Cepila:2018zky,Luszczak:2017dwf,Goncalves:2020vdp,Mantysaari:2017dwh,Cepila:2017nef,Guzey:2018tlk,Henkels:2020kju}, since it can offer additional information on the structure of nuclei. Indeed, similar to the Good–Walker approach~\cite{Good:1960ba} to soft diffraction, in pQCD the ratio of inelastic and elastic hard diffraction at $t=0$ is expressed through the variance of the gluon field in the target~\cite{Frankfurt:2008vi} and is sensitive to onset of black disk regime in which this ratio tends to zero at  high energies. In particular, the ratio was predicted to drop at energies accessible at the LHC~\cite{Cepila:2016uku}.

The ALICE, LHCb, and CMS collaborations have conducted comprehensive studies on $J/\psi$ photoproduction in UPCs~\cite{ALICE:2012yye, ALICE:2013wjo, ALICE:2019tqa, ALICE:2021gpt,ALICE:2021tyx,LHCb:2022ahs,LHCb:2021bfl, CMS:2016itn}.
The ALICE experiment reported a large suppression of the coherent $J/\psi$ cross section at midrapidity as compared to the impulse approximation~\cite{ALICE:2013wjo, ALICE:2021gpt} in reasonable agreement with LTA predictions~\cite{Rebyakova:2011vf, Guzey:2016piu}, with the hot-spot model~\cite{Cepila:2017nef} and the color dipole approach coupled to the solutions of the impact-parameter dependent Balitsky-Kovchegov equation~\cite{Bendova:2020hbb}. At the same time, the ALICE measurement of the incoherent $J/\psi$ photoproduction cross section at central rapidities indicated weaker suppression than predicted by the LTA model~\cite{ALICE:2013wjo}. Later it was shown~\cite{Guzey:2018tlk} that such a discrepancy can be explained if, in addition to incoherent quasielastic events where target nucleons stay intact, one takes into account nucleon dissociative contributions.

On the other hand, ALICE measurements on the coherent $J/\psi$ photoproduction at forward rapidity~\cite{ALICE:2019tqa} appeared to be in tension with LTA and color-dipole model predictions. However, interpretation of the observed discrepancy is not straightforward. In general, extraction of photonuclear cross sections and nuclear gluon densities from UPC measurements at forward rapidities is hampered by the ambiguity caused by the dual role of the colliding nuclei: both nuclei serve as a source of photons and as a target. In order to decouple two contributions to the UPC cross sections, we proposed to study vector meson photoproduction accompanied by additional photon exchanges between colliding ions~\cite{Rebyakova:2011vf, Guzey:2013jaa}. The additional photon exchanges may result in electromagnetic (EM) excitation of one or both nuclei followed by emission of neutrons. The lifetime of the EM excited nucleus is much longer than the time scale of the hard photoproduction process. Hence, both processes can be considered as independent. However the requirement of additional photon exchanges effectively reduces the range of impact parameters and modifies the flux of photons participating in the photoproduction process~\cite{Baltz:2002pp}.  Therefore two independent UPC measurements, with and without neutron emission, can be used to disentangle photonuclear cross sections\footnote{An alternative method based on coherent $J/\psi$ measurements in peripheral and ultra-peripheral collisions was proposed in~\cite{Contreras:2016pkc}.}.

Emission of neutrons due to electromagnetic dissociation of colliding nuclei can be experimentally measured with zero-degree calorimeters (ZDCs) placed at small angles on both sides from the interaction point. UPC events are usually classified into four categories: 0n0n -- no neutrons on both sides; 0nXn and Xn0n -- no neutrons on one side and presence of neutrons on the other side; XnXn -- presence of neutrons on both sides. Calculations of photon fluxes corresponding to these event classes have passed solid experimental tests. First, models of electromagnetic dissociation~\cite{Pshenichnov:2011zz, Broz:2019kpl} have been validated by ALICE measurements of electromagnetic dissociation cross sections in Pb--Pb UPCs~\cite{ALICE:2012aa, ALICE:2022iqi}. Second, the factorization of the coherent photoproduction process and additional photon exchanges has been confirmed by coherent $\rho^{0}$ photoproduction measurements performed by ALICE in different neutron emission classes (0n0n, 0nXn, XnXn) at midrapidity~\cite{ALICE:2021jnv, ALICE:2020ugp, ALICE:2015nbw}.

In the case of incoherent vector meson photoproduction, photons interact with individual nucleons of the target nuclei so one should consider other sources of neutrons in addition to EM dissociation~\cite{Strikman:2005ze}. First, target nucleons can be knocked out from the nucleus and be detected by ZDCs. Second, the incoherent interaction results in the excitation of the target nucleus. If the excitation energy is higher than the energy of neutron separation, neutrons are evaporated and detected by ZDCs with high probability.

New ALICE and STAR results on the coherent and incoherent $J/\psi$ photoproduction in different neutron emission classes are expected to be released soon. We present the LTA predictions for these processes focusing on different mechanisms of the neutron production in coherent and incoherent photoproduction processes.

\section{Photoproduction of $J/\psi$ on nuclei in the leading twist approximation}

The $J/\psi$ photoproduction on nuclear targets is subject to nuclear modifications which depend on the photon-nucleon centre-of-mass energy $W_{\gamma N}$ and can be accounted for by the suppression factor $S(x,\mu^2)$\footnote{Generally, the suppression factor also depends on the squared transverse momentum $t^2_{\perp}$ but in the case of $J/\psi$ photoproduction at relatively small $t$ one can neglect the influence of this effect on the nuclear suppression factor~\cite{Guzey:2016qwo}.}. Then the coherent and incoherent $J/\psi$ photonuclear cross sections can be written in the factorized form as a product of the corresponding cross sections in the impulse approximation (IA) and the suppression factors $S(x,\mu^2)$:
\begin{equation}
{{\rm d^2}\sigma_{\gamma A\rightarrow J/\psi X}(W_{\gamma N},t)\over  {\rm d}y{\rm d}t} =
\frac{{\rm d^2}\sigma_{\gamma A\rightarrow J/\psi X}^{\rm IA}
(W_{\gamma N},t)}{{\rm d}y{\rm d}t}S(x,\mu^2)\,.
\label{eq:sigma_pQCD_2}
\end{equation}
Here $X$ is the final state of the target: $X=A$ in the coherent photoproduction; $X=A^\prime +N$ in the incoherent quasielastic and $X=A^\prime+Y$, where $Y$ is the final state of the incoherent process including the process of diffractive dissociation of the target nucleon\footnote{Inelastic dissociation corresponds to production of several low-energy particles which could interact on the way out of the nucleus and likely would generate several neutrons~\cite{Larionov:2018igy}.}. 

One needs to estimate with high precision coherent and incoherent photoproduction cross sections in the impulse approximation. The coherent cross section is given by the following expression: 
\begin{equation}
 {{{\rm d}\sigma^{\rm  IA}_{\gamma A\rightarrow J/\psi A}(W_{\gamma N},t)}\over {{\rm d}t}}=
 {{{\rm d}\sigma_{\gamma N\rightarrow J/\psi N }(W_{\gamma N},t=0)\over {\rm d}t} \Phi_A(t)} \,.
\label{eq:sigmaIA}
\end{equation}
The forward $\gamma N\rightarrow J/\psi N$ cross section can be calculated in pQCD~\cite{Ryskin:1992ui,Jones:2013pga,Guzey:2013qza} or extracted from experimental data~\cite{H1:2013okq}. In our calculations, we use the following parametrization for the forward cross section~\cite{Guzey:2013xba}: 
\begin{equation}
  \frac{{\rm d}\sigma_{\gamma N \to J/\psi N}(W_{\gamma N},t=0)}{{\rm d} t}=
  C_0 \left[1-\frac{(M_{J/\psi}+m_N)^2}{W_{\gamma N}^2}\right]^{1.5}
  \left(W_{\gamma N}^2/W_0^2\right)^{\delta} \,,
\label{eq:sigma_p}
\end{equation}
where $C_0=342 \pm 8$ nb/GeV$^2$, $\delta=0.40 \pm 0.01$, $W_0=100$ GeV, $M_{J/\psi}$ and $m_N$ are $J/\psi$ and nucleon masses, respectively. This parametrization provides a good description of the available data on the $\gamma p \to J/\psi p$ process measured in fixed-target experiments~\cite{Camerini:1975cy, Binkley:1981kv, EuropeanMuon:1979nky} and at HERA~\cite{ZEUS:2002wfj, H1:2005dtp, H1:2013okq} and is in good agreement with cross sections extracted by ALICE and LHCb from the data on exclusive $J/\psi$ photoproduction off protons in pp and p--Pb UPCs at the LHC~\cite{ALICE:2014eof,ALICE:2018oyo,LHCb:2014acg,LHCb:2018rcm}.

The $t$ dependence of the coherent photonuclear cross section, $\Phi_A (t)$, is driven by the squared nuclear form factor: $\Phi_A (t)=| AF_A(t) |^2$. The $t$-integrated cross section can be obtained as~\footnote{The presence of shadowing changes the shape of the $t$ dependence and may result in a 10\%-tish correction to the $t$-integrated cross section~\cite{Guzey:2016qwo}.}:
\begin{equation}
\Phi_A(t_{\rm min})=\int \limits_{t_{\rm min}}^{\infty} dt 
{\left|AF_A (t)
\right|
}^2 \,,
\label{fi}
\end{equation}
where $t_{\rm min}=-p_{z,{\rm min}}^2=-[M^2_{J/\psi}/(4\omega \gamma _L)]^2$ is determined by the minimal longitudinal momentum transfer $p_{z,{\rm min}}$ characterizing the coherence length which becomes important in the low energy domain. Here $\gamma_L$ is the nucleus Lorentz factor, $\omega$ is the photon energy.

For the quasielastic incoherent $J/\psi$ photoproduction on nucleus the $t$ dependence can be approximated by the expression $\Phi^{qel}_A (t)=A \left [1-{\mid F_{A}(t)\mid}^2\right]  \exp[B^{el}_{J/\psi N}(W_{\gamma N}) t]$ with the slope parameter $B^{el}_{J/\psi N}(W_{\gamma N})=4.5+0.5\ln{W_{\gamma N}\over 90}$ extracted from H1 measurements~\cite{H1:2013okq}.

Finally, for the incoherent process with diffractive dissociation of the target nucleon (Ntd), we use the following parametrization of the cross section provided by the H1 experiment~\cite{H1:2013okq}:
\begin{equation}
{\frac{{\rm d}\sigma^{\rm IA}_{\gamma N \to J/\psi Y}(W_{\gamma N},t)}{{\rm d}t}}=
{(66\pm 10){\biggl[{W_{\gamma N}\over 90}\biggr ]}^{0.42}}\Phi^{\rm Ntd}_A (t)\, ,
\end{equation}
where $\Phi^{\rm Ntd}_{A}(t)=A[1+{{1.79\mid t_\perp \mid} \over {3.58}}]^{-3.58}$.
 
In the coherent case, the nuclear suppression factor is characterized by the squared ratio of the nuclear $G_A(x,\mu^2)$ and nucleon $G_N(x,\mu^2)$ gluon density functions:
\begin{equation}
S_{\rm coh}(x,\mu^2)= R^2(x,\mu^2)=\left[\frac{G_A(x,\mu^2)}{AG_N(x,\mu^2)}\right]^2\,.
\end{equation} 

The quantity $S(x,\mu ^2)$ can  be calculated within the LTA model of parton shadowing, which combines geometry of the Glauber  theory of the multiple hadron-nucleus scattering~\cite{Glauber:1959}, the Gribov connection of shadowing and diffraction~\cite{Gribov:1968jf} (see for details \cite{Gribov:2009zz}) and QCD factorization theorems for inclusive and diffractive parton distributions~\cite{Frankfurt:2011cs}, in terms of the effective cross sections $\sigma_{2}(x,\mu^2)$ and $\sigma_{3}(x,\mu^2)$:
\begin{equation}
  S_{\rm coh}(x,\mu^2)= {\left[{1-\frac {\sigma_{2}(x,\mu^2)} {\sigma_{3}(x,\mu^2)}}
 \biggl (1-\frac {2\int {\rm d}^2 {\bf b}  [1-e^{-{{\sigma_{3}(x,\mu^2)}\over {2}} T_A({\bf b})}]}
 {A \sigma_{3}(x,\mu^2)}\biggr )\right ]}^2 \, ,
\label{eq:sfact_coh_2}
\end{equation}
where $T_A({\bf b})=\int {\rm d}z\, \rho_A({\bf b},z)$ is the nuclear thickness function. The effective cross section $\sigma_{2}(x,\mu^2)$ describes rescattering of the quark-gluon configurations in the photon wave function on two nucleons and is determined by the parameters of diffractive dissociation cross section on the nucleon and diffractive parton distributions. The effective cross section  $\sigma_{3}(x,\mu^2)$ presents the soft interaction of the configurations of different sizes in the photon wave function with three and more nuclear nucleons. This quantity can't be calculated in pQCD and is estimated in the LTA using the Good-Walker formalism \cite{Good:1960ba} of the cross section fluctuations resulting in two models which lead to a weaker (WS) and to a stronger (SS) gluon shadowing in the LTA\footnote{In the LTA, only one parton is involved in the hard interaction in the {\it in} and {\it out} states.}. The model of stronger shadowing is based on assumption that $\sigma_{3}(x,\mu^2)\approx \sigma_{2}(x,\mu^2)$ that leads to the limit $S_{\rm coh}(x,\mu^2)\approx \frac {\sigma^{tot}_{A}(\sigma_{2})} {A\sigma_{2}(x,\mu ^2)}$. The second model resulting in weaker shadowing assumes that configurations of different sizes in the photon wave function interact  with three and more nuclear nucleons with cross sections characterized by the distribution $P(\sigma)$ similar to the $P_{\pi}(\sigma)$ of pion.  Such modeling leads to rather large uncertainties in the LTA predictions at $x<10^{-3}$ though both models predict a strong shadowing.

The LTA framework can be also used for the calculation of the suppression factor in the case of incoherent $J/\psi$ photoproduction~\cite{Guzey:2013jaa}:
\begin{eqnarray}
S_{\rm incoh}(x,\mu^2)={1\over {A}}
\int {\rm d}^2 {\bf b}\, T_A({\bf b}) \left[1-\frac{\sigma_{2}(x,\mu^2)}{\sigma_{3}(x,\mu^2)}
\biggl [{1-e^{-{{\sigma_{3}(x,\mu^2)}\over {2}} T_A({\bf b})}\biggr]} \right]^2\, .
\label{eq:sfact_incoh_2}
\end{eqnarray}

\begin{figure}[p!]
\centering
\epsfig{file=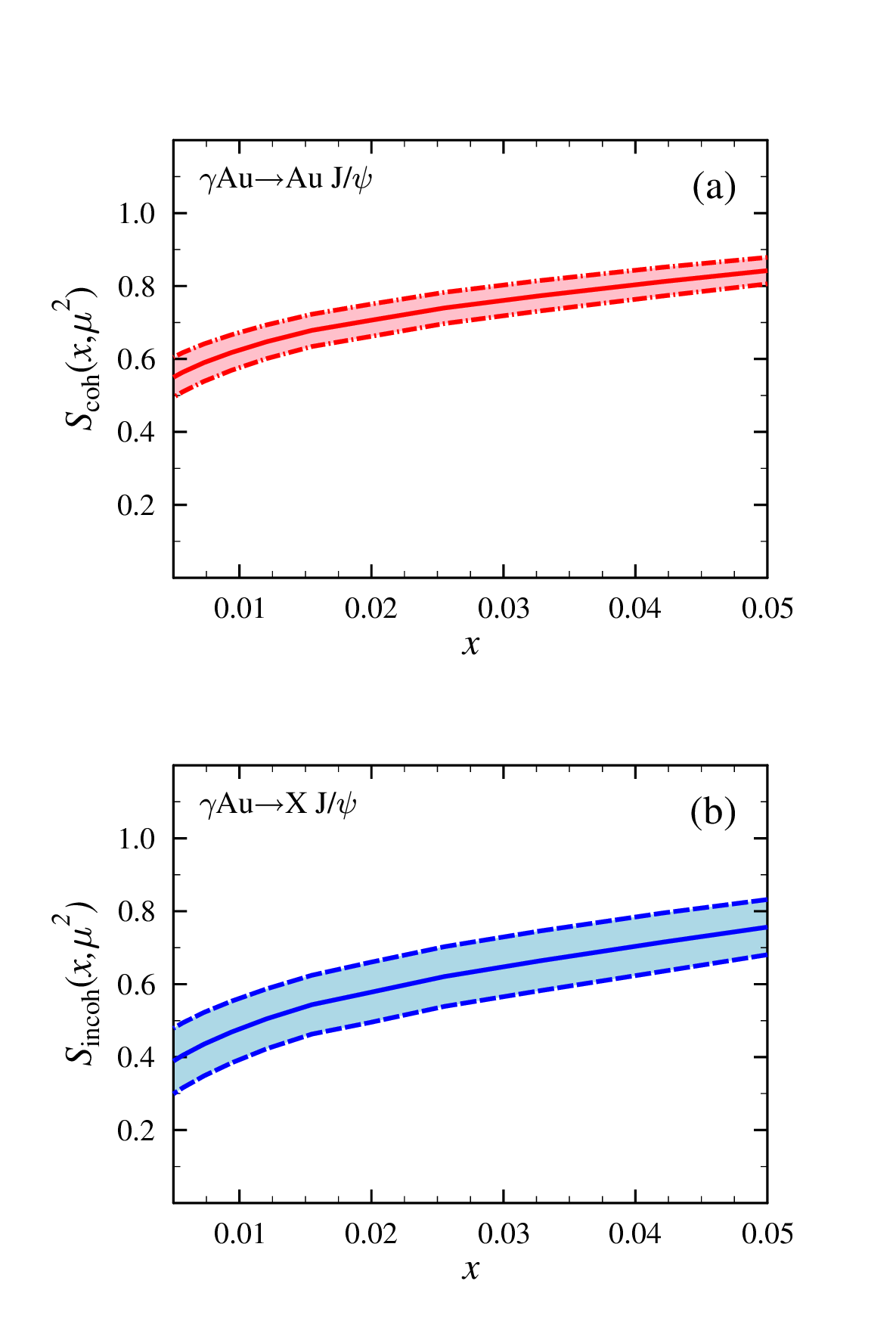,scale=0.7}
\caption{The nuclear suppression factor $S(x,{\mu}^2)$ predicted in the LTA model for coherent (a) and incoherent (b) photoproduction in Au--Au UPCs at $\sqrt{s_{\rm NN}} =200$ GeV. The range $0.005\leq x\leq 0.05$ corresponds to the rapidity interval $|y|< 1$ studied in the STAR experiment.}
\label{fig:shadrhic}
\end{figure}

\begin{figure}[p!]
\centering
\epsfig{file=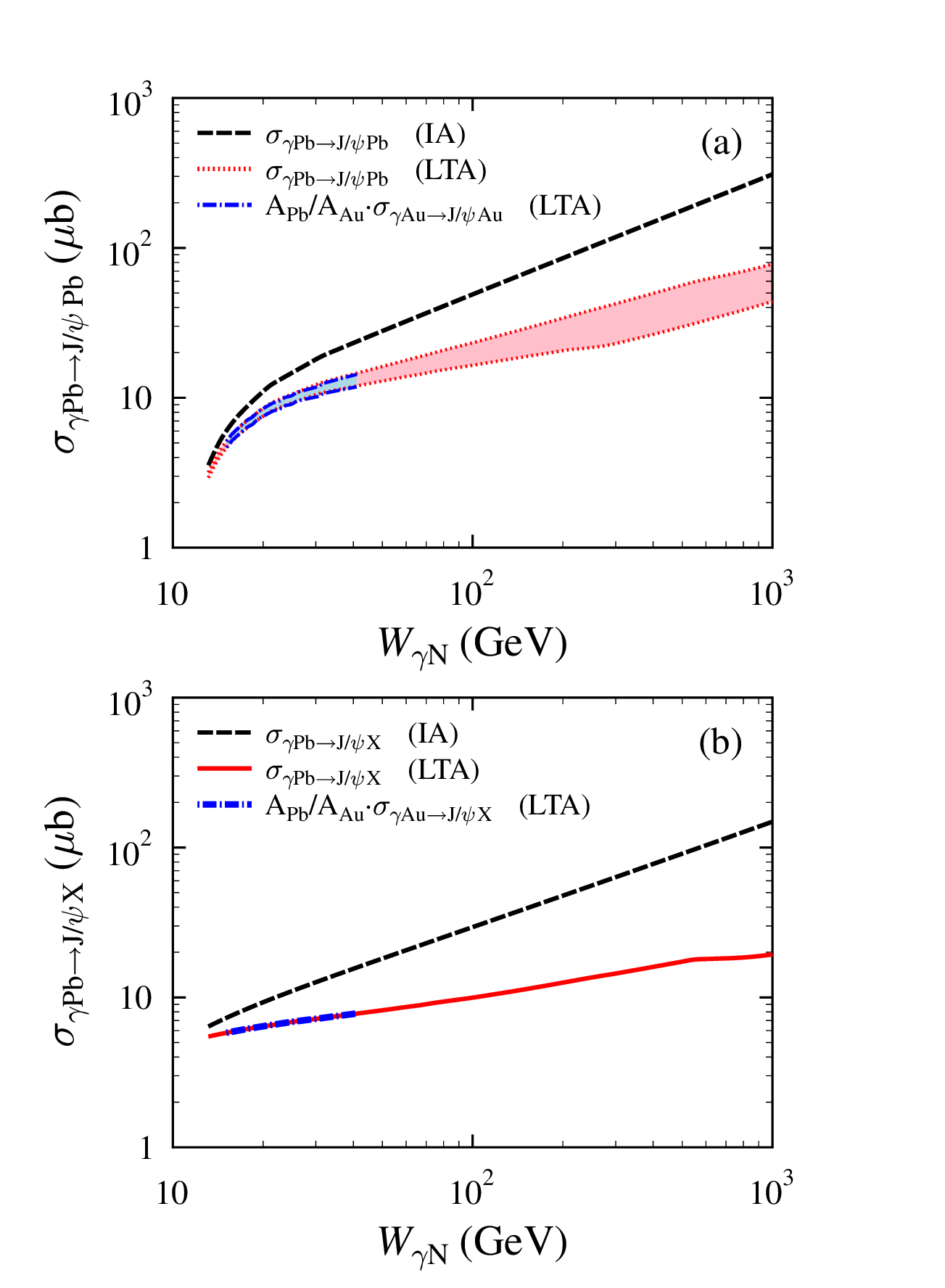,scale=0.7}
\caption{Coherent (a) and incoherent (b) photonuclear cross sections $\gamma {\rm Pb}\rightarrow {\rm Pb} J/\psi$ as a function of photon-nucleon centre-of-mass energy $W_{\gamma N}$ in the impulse approximation (IA) compared to the LTA model predictions. The shaded area indicates the LTA model uncertainty. Light blue strip shows the $\gamma {\rm Au}\rightarrow {\rm Au} J/\psi$ cross section that can be measured in Au--Au UPCs at RHIC scaled by the ratio of the numbers of nucleons in Pb and Au. The bottom panel shows the energy dependence for the sum of the incoherent quasielastic and nucleon dissociative photoproduction cross sections (upper limit).}
\label{fig:endepga}
\end{figure}

Nuclear suppression factors obtained in the LTA for the coherent and incoherent $J/\psi$ photoproduction are shown in Fig.~\ref{fig:shadrhic} as a function of $x$ in the range accessible with the STAR detector at RHIC. Fig.~\ref{fig:endepga} illustrates the $W_{\gamma N}$ dependence of the coherent and incoherent\footnote{Note that we give here  the upper limit (integral over $-t_{\perp}$ up to $\infty$) for the energy dependence of the incoherent cross section.}  photonuclear cross sections, obtained in the impulse approximation and within the LTA framework, indicating the range of energies that can be studied with the STAR detector at RHIC and with the ALICE detector at the LHC. It is worth noting that the study of the coherent photoproduction cross section in the range of rapidities of the STAR detector can be useful for understanding of the discrepancy between ALICE measurements at forward rapidity and LTA predictions.

\begin{figure}[p!]
\centering
\epsfig{file=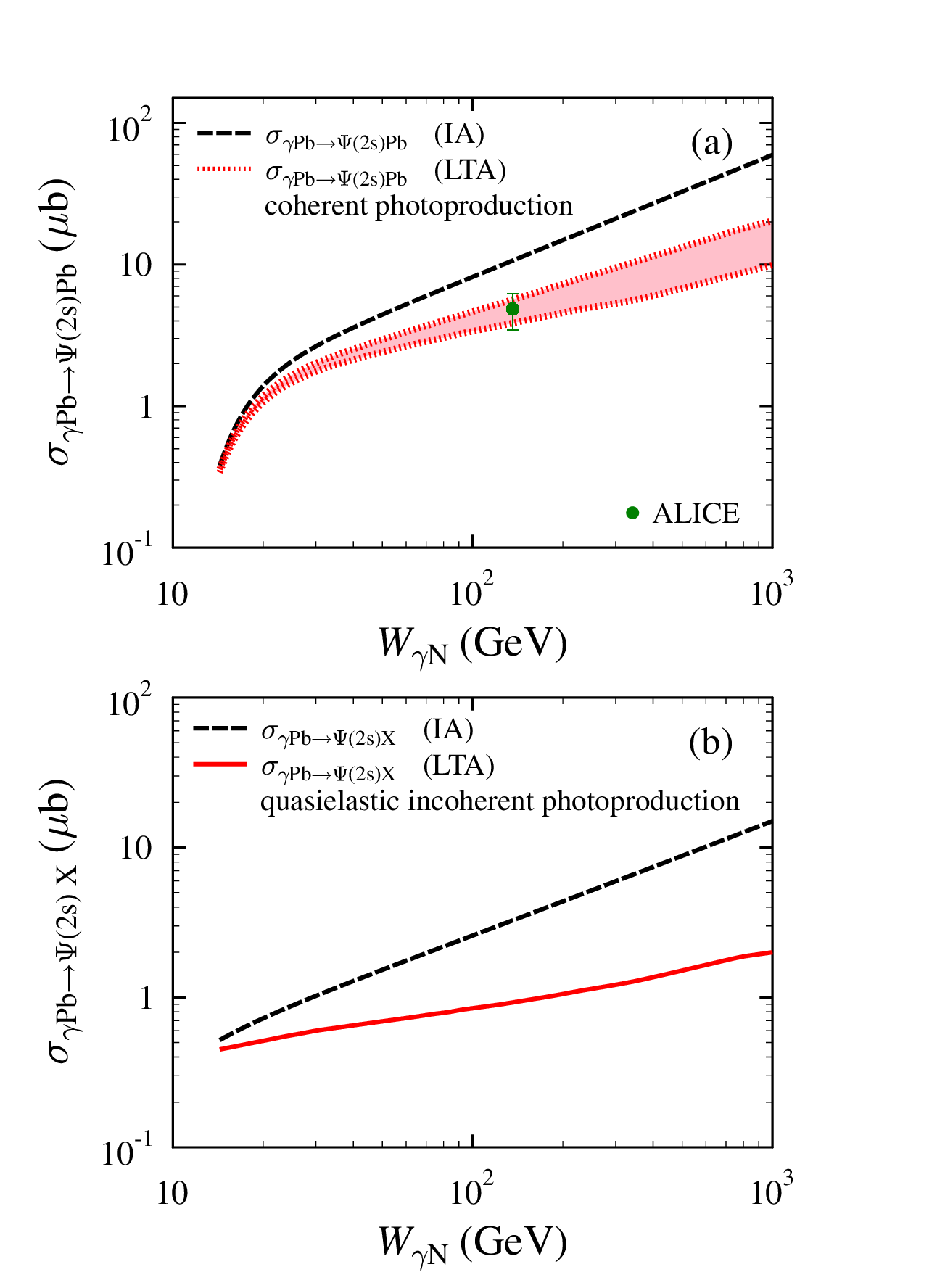,scale=0.7}
\caption{Coherent (a) and incoherent (b) photonuclear cross sections $\gamma {\rm Pb}\rightarrow {\rm Pb} \psi(2S)$ as a function of photon-nucleon centre-of-mass energy $W_{\gamma N}$ in the impulse approximation (IA) compared to the LTA model predictions. The coherent $\psi(2S)$ cross section extracted from the ALICE measurement~\cite{ALICE:2021gpt} is shown with green marker.
The shaded area in the coherent cross section indicates the LTA model uncertainty. Nucleon dissociative component is not taken into account in the incoherent cross section. 
}
\label{fig:psi2s}
\end{figure}

The same approach with minor modifications can be applied to the coherent and incoherent photoproduction of $\psi(2S)$ mesons. For completeness, fig.~\ref{fig:psi2s} presents LTA predictions for the energy dependence of coherent and incoherent $\psi(2S)$ photoproduction cross sections based on existing measurements of $\psi(2S)$ photoproduction off protons at HERA~\cite{H1:2002yab,ZEUS:2022sxn}. The coherent photonuclear cross section is in good agreement with the ALICE measurement~\cite{ALICE:2021gpt}.
An alternative treatment of $\psi(2S)$ photoproduction within the color-dipole approach can be found in~\cite{Henkels:2020kju}.

Within the LTA framework, the $t$-integrated coherent and incoherent cross sections of $J/\psi$ photoproduction on nucleus are determined by almost the same quantities: the nuclear density $\rho_{A}(r)$ and the nuclear form factor are known from experiments on electron and proton elastic scattering on nuclei; the forward $\gamma N\rightarrow J/\psi N$ cross section and the slope parameter of the $t$ dependence have been measured in a wide range of energies at HERA and in proton-proton and proton-nucleus UPCs at the LHC. Hence, $J/\psi$ photoproduction measurements in nucleus-nucleus UPCs allows one to extract $S_{\rm coh}(x,\mu^2)$ and $S_{\rm incoh}(x,\mu^2)$ values and determine both $\sigma_{2}(x,\mu^2)$ and $\sigma_{3}(x,\mu^2)$ from Eqs.~\ref{eq:sfact_coh_2} and \ref{eq:sfact_incoh_2} at the corresponding $x$ thus improving  accuracy of the LTA predictions and testing the self-consistency of the LTA model.

\section {$J/\psi$ photoproduction cross sections in UPCs}

Both colliding nuclei in the UPCs serve as sources of quasi-real photons and as targets. Hence, the cross section of $J/\psi$ photoproduction in symmetric collisions of ions with per-nucleon energy $E_N$  is given by the sum of two terms for each class of events (coherent, incoherent quasielastic, incoherent with nucleon dissociation, all with or without electromagnetic excitation by additional photon exchange):
\begin{eqnarray}
\frac{{\rm d}\sigma_{AA \to J/\psi A^* A^*}(y)}{{\rm d}y{\rm d}t}=
N_{\gamma/A}(W^{(+)}_{\gamma N})
{{{\rm d}\sigma_{\gamma A \to J/\psi A^*}(W^{(+)}_{\gamma N})}
\over {{\rm d}y {\rm d}t}}
+ N_{\gamma/A}(W^{(-)}_{\gamma N}) 
\frac{{\rm d}\sigma_{\gamma A \to J/\psi A^*}(W^{(-)}_{\gamma N})}{{\rm d}y {\rm d}t} 
\label{eq:sigma_AA}
\end{eqnarray}

Here $y$ is the rapidity of $J/\psi$ defined with respect to the right-moving nucleus emitting the photon (the first term in Eq.~\ref{eq:sigma_AA}) and related to the invariant photon-nucleon energy 
$W^{(\pm )}_{\gamma N} = \sqrt{2 E_N M_{J/\psi}}\,e^{\pm y/2}$ with the plus sign in the exponent. The  second term corresponds to the left-moving photon source and the minus sign in the expression for $W_{\gamma N}$. Symbol $A^*$ is the final state of colliding nuclei, $N_{\gamma/A}(W_{\gamma N})$ is the photon flux, and $\sigma_{\gamma A \to J/\psi A^*}(W_{\gamma N})$ is the $J/\psi$ photoproduction cross section on nuclear target.

The flux of photons with energy $\omega =\frac {M_{J/\psi}} {2}e^{\pm y}$ can be calculated as the 
convolution of two terms 
\begin{equation}
 N_{\gamma /A}(\omega )= \int \limits_{2R_A}^{\infty} d^2b\,  
 N_{\gamma /A}(\omega ,\vec b) P_{i}(\vec b).
 \label{flux}
\end{equation}
Probabilities $P_{i}(\vec b)$ select classes of events with or without additional photon exchanges with subsequent neutron emission which we calculate based on parametrization given in~\cite{Baltz:2007kq}. This approach is also used in STARLIGHT~\cite{Klein:2016yzr} and nOOn~\cite{Broz:2019kpl} Monte--Carlo generators of UPC events. $N_{\gamma /A}(\omega ,\vec b)$ is the flux of equivalent photons produced by a point-like particle with the electric charge $Z$: 
\begin{eqnarray}
 N_{\gamma /A}(\omega,\vec b )= \frac {Z^2\alpha_{\rm e.m.}} {{\pi }^2} \,
 {\zeta ^2\over b^2} \left[ K^2_1(\zeta ) +\frac {1} {\gamma^2_L} K^2_0(\zeta )\right]
\label{flux2}
\end{eqnarray}
where $\alpha_{\rm e.m.}$ is the fine-structure constant, $\vec b$ is the impact parameter between colliding nuclei,  $K_0(\zeta )$ and $K_1(\zeta )$ are modified Bessel functions with argument $\zeta =\frac {b\omega} {\gamma_L}$ depending on the photon energy and impact parameter. 

Strong interactions between colliding nuclei in UPCs should be absent. Experimentally this is achieved by requiring no signal in the detector except for two tracks from dilepton $J/\psi$ decays, while in calculations the absence of strong interactions is taken into account with the condition on impact parameter $b>2R_A$ which can be additionally strengthened by convolution of the photon flux with a factor suppressing probability of strong interactions of the overlapping nuclei.

\section{Results}

\begin{figure}[p!]
\centering
\epsfig{file=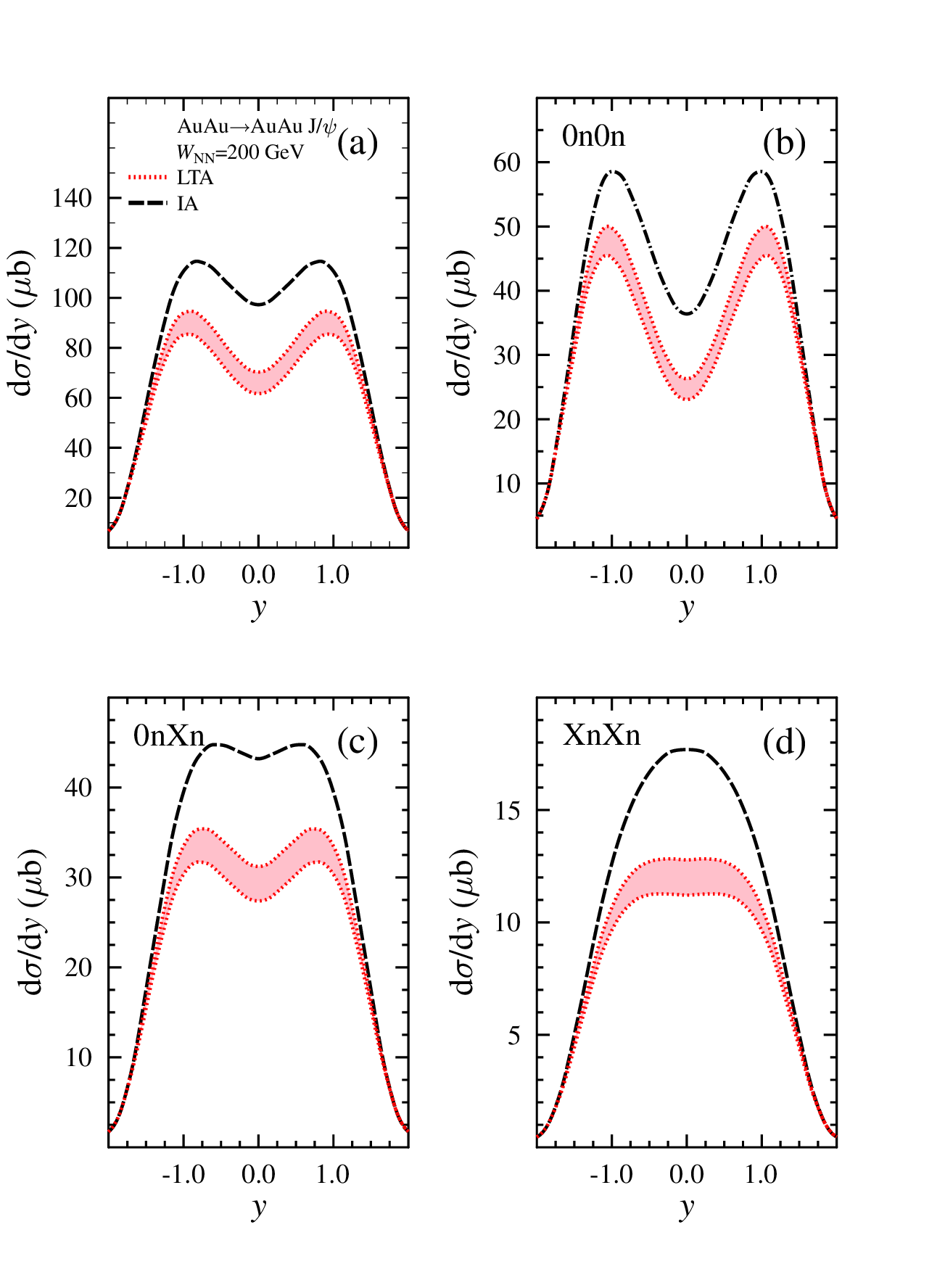,scale=0.7}
\caption{Rapidity distribution for the coherent $J/\psi$ photoproduction in Au--Au UPCs at $\sqrt{s_{\rm NN}}=200$ GeV in different neutron emission classes: (a) -- total, (b) -- 0n0n, (c) -- 0nXn, and (d) -- XnXn. Each panel contains two curves: dashed black curve shows coherent $J/\psi$ photoproduction cross sections calculated in the impulse approximation (IA); dotted red -- coherent photoproduction cross section calculated in the LTA model with the shaded area indicating the LTA model uncertainty.} 
\label{fig:raprhic}
\end{figure}

\begin{figure}[p!]
\centering
\epsfig{file=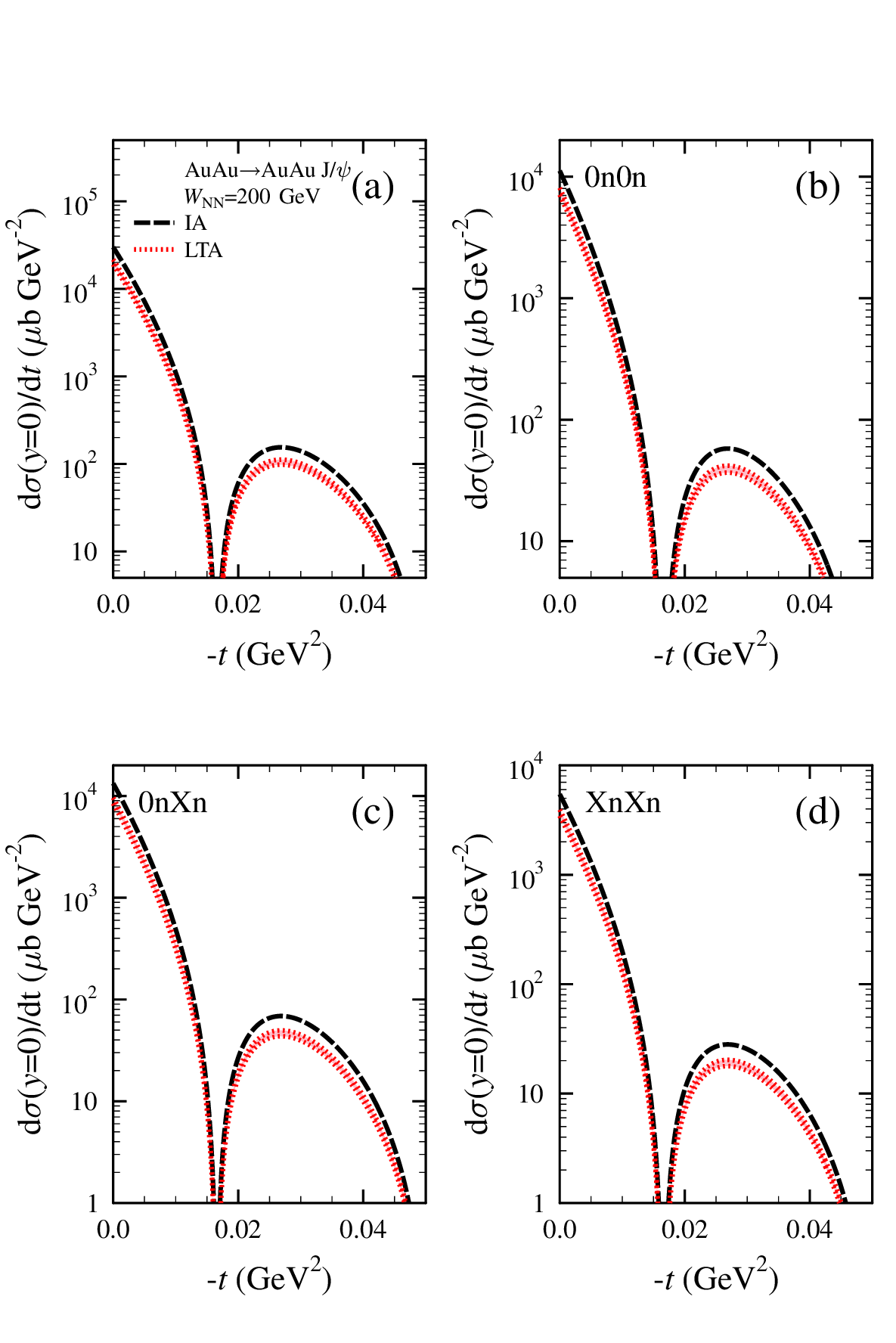,scale=0.7}
\caption{Predictions for the $t$ distribution of the coherent $J/\psi$ photoproduction in Au--Au UPCs  at $\sqrt{s_{\rm NN}}=200$ GeV in different neutron emission classes: (a) -- total, (b) -- 0n0n, (c) -- 0nXn, and (d) -- XnXn. Each panel contains two curves: dashed black curve represents coherent photoproduction cross sections calculated in the impulse approximation; dotted red -- coherent photoproduction cross section in the LTA model.}
\label{fig:dtcohrhic}
\end{figure}

Using the inputs described above, we can make predictions  for coherent and incoherent $J/\psi$ photoproduction cross sections in ultraperipheral Au--Au collisions at $\sqrt{s_{\rm NN}}=200$ GeV at RHIC and incoherent cross sections for Pb--Pb collisions at the LHC\footnote{Predictions for coherent $J/\psi$ photoproduction cross sections in  Pb--Pb UPCs at the LHC were provided in~\cite{Guzey:2016piu}.}. The $t$-integrated rapidity distributions for coherent cross sections in Au--Au UPCs at RHIC  calculated in the LTA are compared to the impulse approximation in Fig.~\ref{fig:raprhic} for different neutron emission classes. The $t$ dependence of the coherent photoproduction in UPCs at $y=0$ is shown in Fig.~\ref{fig:dtcohrhic}. It should be possible to compare these distributions with future STAR measurements in the range $|y|<1$ corresponding to the energy range $15 < W_{\gamma N} <41$~GeV. 

The most interesting information which can be obtained from these data is the nuclear suppression factors which characterize gluon shadowing in Au nuclei. Note that the values of $S_{\rm coh}(x,\mu^2)$ can be directly found from the data only at $y=0$ where both contributions in Eq.~\ref{eq:sigma_AA} coincide. From Fig.~\ref{fig:shadrhic} one can find that, at $y=0$ corresponding to $W_{\gamma N}\approx 25$ GeV and $x\approx 0.015$, the nuclear gluon shadowing in coherent $J/\psi$  photoproduction on Au is $R(x\approx 0.015,{\mu ^2}=3\, {\rm GeV}^2)\approx (0.8\div 0.85) $. To determine the shadowing factor  from the experimental coherent rapidity distribution at other values of $x$, one needs to separate photoproduction contributions from lower and higher energies. It can be done with UPC cross section measurements in different neutron emission classes.

We also predict the rapidity distribution and the $t$ dependence of the incoherent photoproduction in Au--Au UPCs which is under study by the STAR collaboration. Recent analysis~\cite{Guzey:2022qvc} of the $t$ dependence of the incoherent $J/\psi$ photoproduction measured by STAR in dAu UPCs revealed that a reasonable description of the data can be obtained under assumption that the main contribution to the cross section is provided by the incoherent quasielastic photoproduction on the deuteron target while the mechanism of photoproduction with diffractive dissociation of the nucleon in deuteron is suppressed by triggering conditions applied in the selection of UPCs.
 
For the rapidity and $t$ dependence of the incoherent $J/\psi$ photoproduction in different neutron emission classes, we consider several neutron emission sources. In contrast to the coherent photoproduction where the transverse momentum transfer is limited by $q_{t}\leq R^{-1}_{A}$, the incoherent quasielastic process is characterised by $q_{t}\leq R^{-1}_N$ and some longitudinal momentum transfer $q^{0}_{l}$. Hence, the projectile interacts with individual nucleons resulting in transition of the target nucleus from ground to excited states or even to the nucleus breakup when $q^0_{l}>150$ MeV/$c$, i.e. when the transferred energy exceeds the minimal nucleon separation energy $E_{\rm S}^{\rm min}\approx 8$ MeV. The target neutrons escape from nuclei and, depending on the energy of UPCs and their transverse momentum, can be captured by ZDCs. In addition, ZDCs will also detect neutrons emitted by the decaying excited residual nucleus if the excitation energy exceeds $E_{\rm S}^{\rm min}$. Also neutrons are emitted by the nucleus excited by additional photon exchanges. 

Measurements of inclusive incoherent quasielastic $J/\psi$ photoproduction in UPCs requiring  only $J/\psi$ detection implies the closure over various final states of the target. Contrary, additional requirements on the presence of neutrons results in the selection of nuclear final states with different excitation energy: probability of neutron emission increases with the excitation energy of the target nucleus. An accurate accounting for these effects requires detailed calculations of the nuclear spectral functions for highly excited nuclear states and partial probabilities of their decays with neutron emission. In principle, such information can be obtained from studies of the nucleon knockout in $A(e,e^\prime p) $ or $A(p,2p)$ reactions at intermediate energies. Almost direct evidence of the neutron evaporation after knockout of a nucleon from the nucleus has been revealed in the data on deep inelastic scattering (DIS) with detection of slow neutrons in the final state~\cite{E665:1995utr}. Monte--Carlo modeling of the decay of nuclear hole states produced in DIS by evaporation of neutrons~\cite{Strikman:1996ak, Strikman:1998cc} provided reasonable description of the momentum dependence of neutron emission. Similar approach was used in~\cite{Strikman:2005ze} to estimate the probability of neutron emission from the target Au nucleus in the process of incoherent photoproduction of $J/\psi$ at RHIC energies. This probability appeared  to be in the range 0.8--0.85, close to the ratio of the number of nucleons with the separation energy $E_{\rm S}>2E^{\rm min}_{\rm S}$ to the total number of nucleons in heavy nuclei. 
In our calculations on the incoherent $J/\psi$ cross section in different neutron emission classes we redistribute the fractions of events in 0n0n, 0nXn and XnXn categories taking into account that 85\% of incoherent events are accompanied by neutron emission.

\begin{figure}[p!]
\centering
\epsfig{file=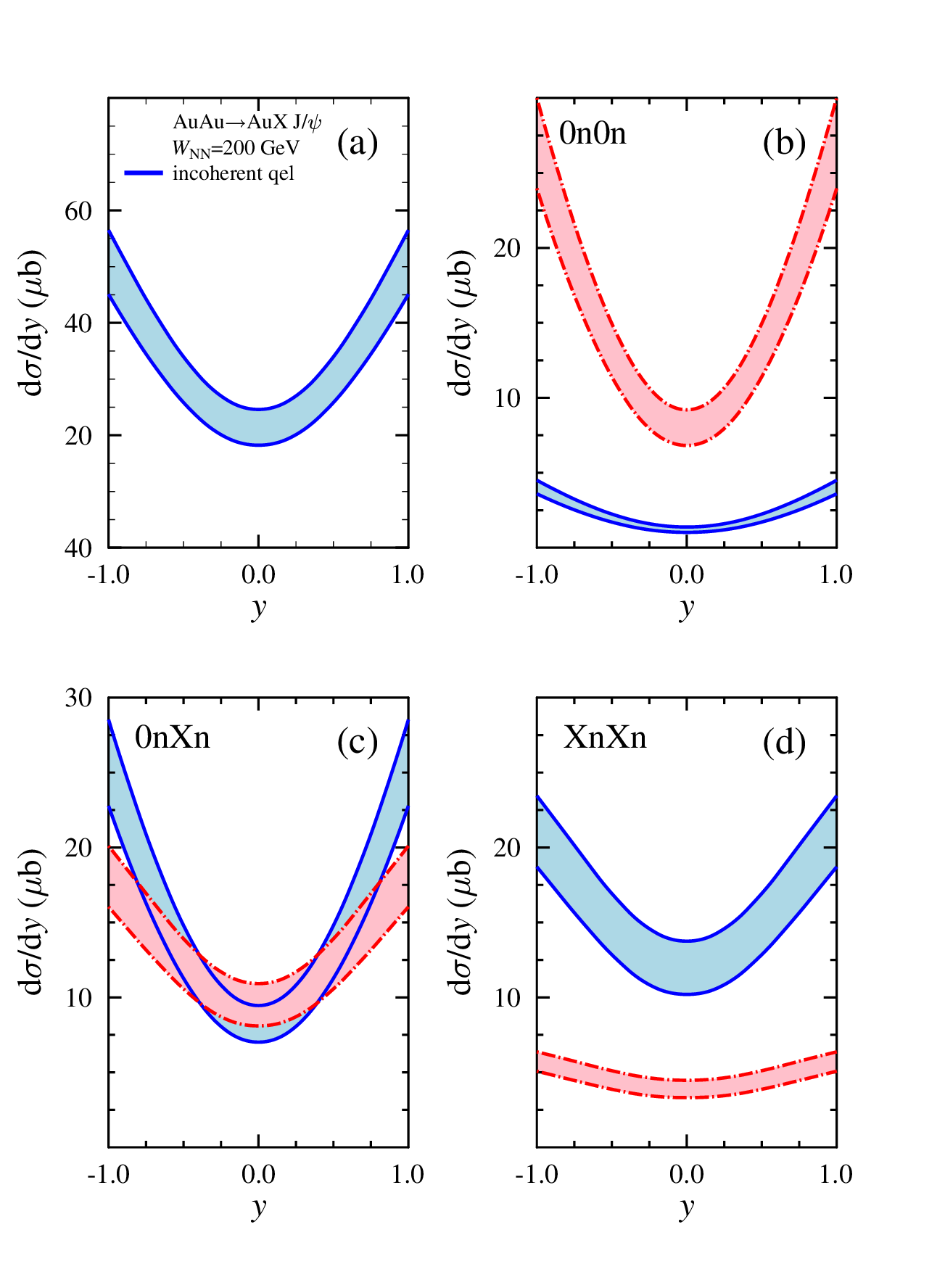,scale=0.7}
\caption{Rapidity distribution for the incoherent quasielastic $J/\psi$ photoproduction in Au--Au UPCs in different neutron emission classes  at $\sqrt{s_{\rm NN}}=200$ GeV: (a) -- total, (b) -- 0n0n, (c) -- 0nXn, and (d) -- XnXn. Each panel contains two curves: dashed red curves show cross sections calculated in the LTA model accounting for EM dissociation only; solid blue lines -- also accounting for neutrons from the decay of the target nucleus excited in the photoproduction process. The shaded bands represent uncertainties of the LTA model.}
\label{fig:rhicinc}
\end{figure}

\begin{figure}[p!]
\centering
\epsfig{file=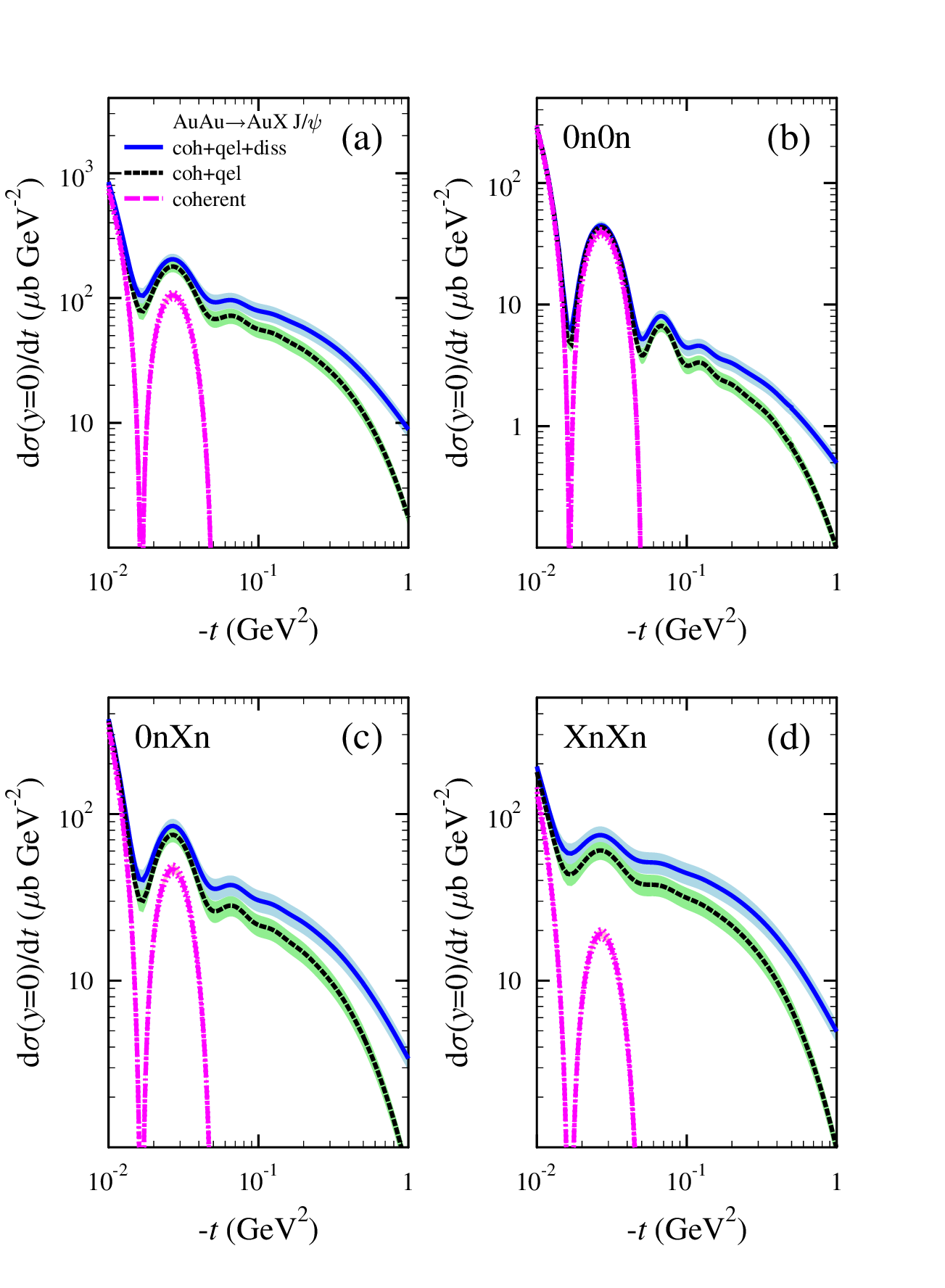,scale=0.7}
\caption{LTA model predictions for $t$ distributions of $J/\psi$ photoproduction in Au--Au UPCs in different neutron emission classes  at $\sqrt{s_{\rm NN}}=200$ GeV: (a) -- total, (b) -- 0n0n, (c) -- 0nXn, and (d) -- XnXn. The dot-dashed magenta line represents the coherent (coh) cross section; short dashed black line -- the sum of coherent and incoherent quasielastic (qel) cross sections; solid blue line -- the sum of coherent, incoherent quasielastic and nucleon dissociative (diss) cross sections.}
\label{fig:incrhic}
\end{figure}

\begin{figure}[p!]
\centering
\epsfig{file=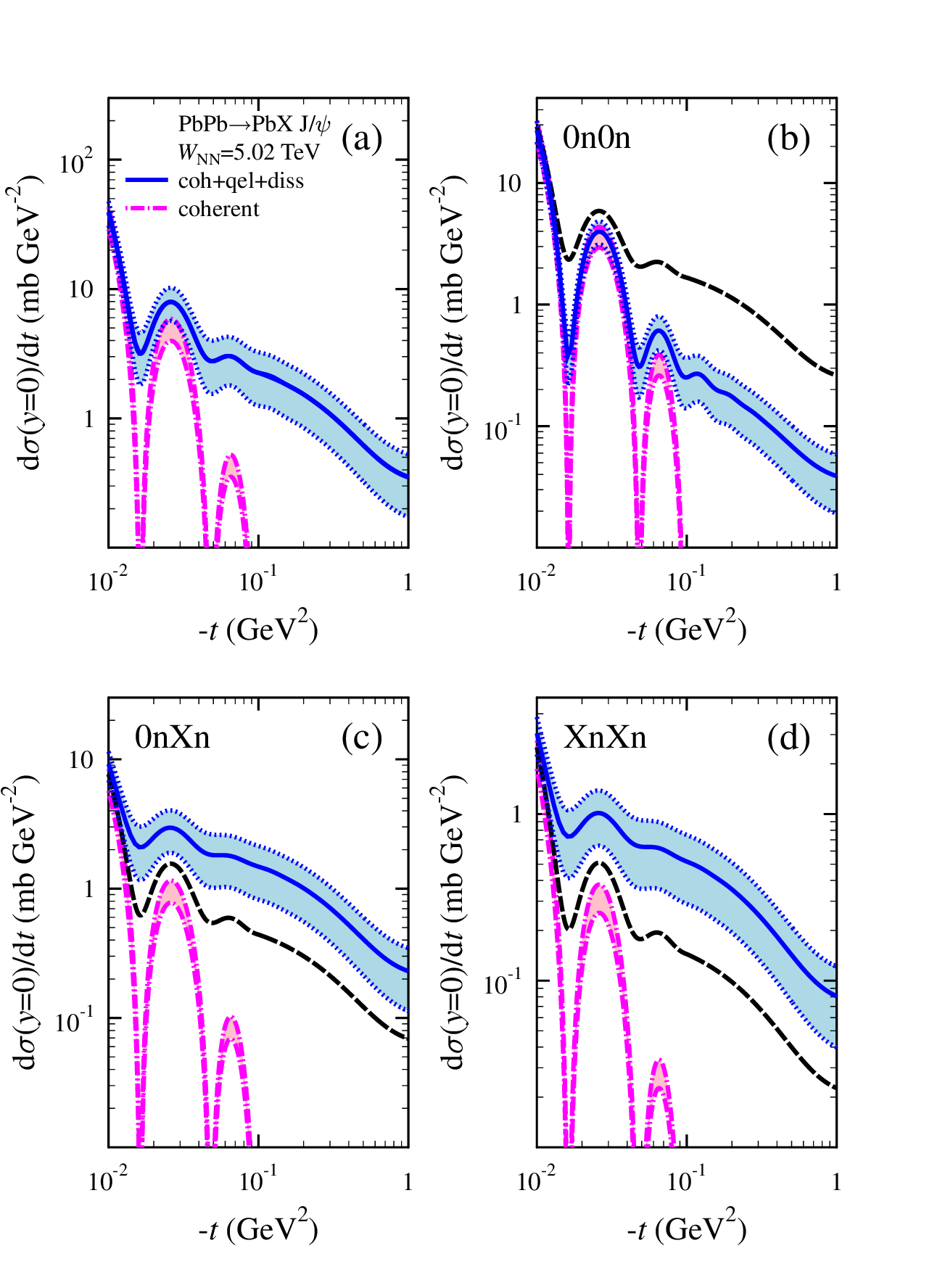,scale=0.7}
\caption{LTA model predictions for $t$ distributions of $J/\psi$ photoproduction in Pb--Pb UPCs in different neutron emission classes  at $\sqrt{s_{\rm NN}}=5020$ GeV: (a) -- total, (b) -- 0n0n, (c) -- 0nXn, and (d) -- XnXn. Each panel contains three sets of curves: dot-dashed magenta line -- coherent photoproduction; dashed black -- cross sections accounting for neutrons emitted due to EM dissociation of nuclei; solid blue -- the sum of coherent, incoherent quasielastic and nucleon dissociative cross sections accounting for neutrons from the EM dissociation and from the decay of the target nucleus excited in the photoproduction process.}
\label{fig:inclhc}
\end{figure}

\begin{figure}[p!]
\centering
\epsfig{file=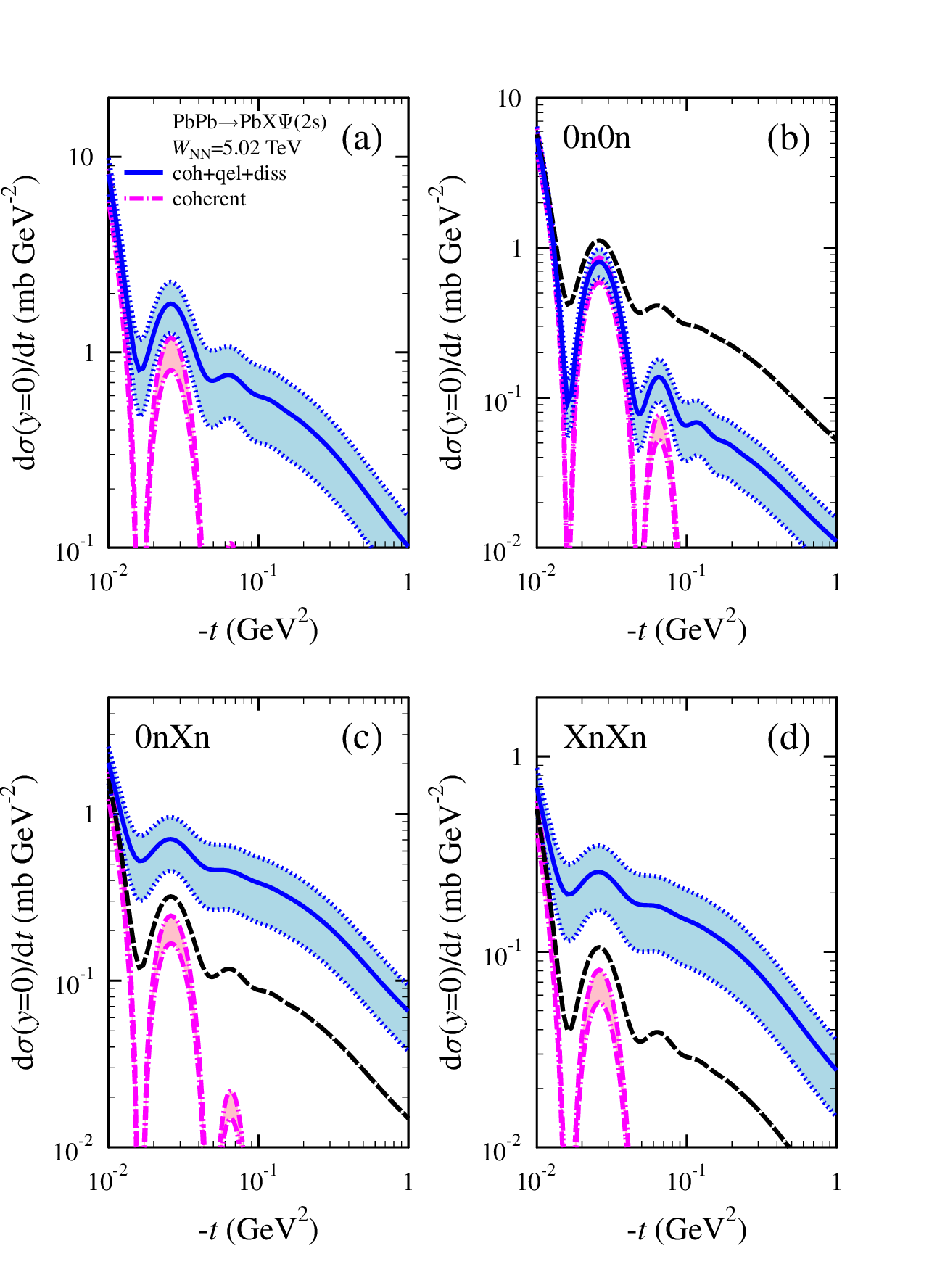,scale=0.7}
\caption{LTA model predictions for $t$ distributions of $\psi(2S)$ photoproduction in Pb--Pb UPCs in different neutron emission classes at $\sqrt{s_{\rm NN}}=5020$ GeV. See caption to Fig.~\ref{fig:inclhc} for details.}
\label{fig:psi2s_t}
\end{figure}

Additional sources of evaporated neutrons have a few important consequences for the studies of incoherent $J/\psi$ photoproduction in UPCs in different neutron emission classes. Accounting for the neutron evaporation by the excited target nucleus results in significant changes of the $J/\psi$ yield in different classes as compared to estimates based only on EM excitations of nuclei due to additional photon exchanges. The main effect is strong suppression of the 0n0n contribution in the total cross section with simultaneous increase of the role of 0nXn and XnXn channels. This is clearly demonstrated in  Fig.~\ref{fig:rhicinc} for the rapidity distributions of the incoherent quasielastic  $J/\psi$ photoproduction in the STAR acceptance in different event classes where red lines correspond to $J/\psi$ cross sections calculated assuming EM dissociation source of neutrons only, while cross sections shown with blue lines also account for neutrons knocked out or evaporated from the excited target nucleus.
Fig.~\ref{fig:incrhic} presents predictions for the $t$ distributions at $y=0$ for incoherent photoproduction in Au--Au UPCs at RHIC where we show the cross sections with  (blue line) and without (black line) accounting for the contribution of the target nucleon dissociation process.  Finally, in  Fig.~\ref{fig:inclhc} we show $t$ distributions for coherent and incoherent $J/\psi$ photoproduction in Pb--Pb UPCs at the LHC demonstrating significant effects from accounting for the neutron evaporation from the excited target nucleus (solid blue line against dashed black one). Similar $t$ distributions for $\psi(2S)$ photoproduction in Pb--Pb UPCs are shown in Fig.~\ref{fig:psi2s_t}.

It was shown in~\cite{Guzey:2013jaa} that measurements in different neutron emission classes can be used to separate two terms in Eq.~\ref{eq:sigma_AA} in the incoherent $J/\psi$ photoproduction in symmetric UPCs. This can be done by studying the 0nXn class using the evident correlation: the direction with zero neutrons tags the nucleus which serves as a source of photons, while the direction with $X$ neutrons identifies the target nucleus.  Selection of UPC events with high-energy $J/\psi$ in the 0n direction accompanied by neutrons in the opposite direction allows one to measure the cross section of incoherent $J/\psi$ photoproduction by high energy photons producing $c \bar c$ pairs which interact with low-$x$ gluons in the nuclear target. Such studies will provide information on low-$x$ gluon shadowing in nuclei which is complimentary to coherent $J/\psi$ photoproduction measurements.

\section{Conclusions}

 Photoproduction measurements in UPCs provide important tests of the shadowing dynamics in hard processes. We presented predictions on the coherent and incoherent $J/\psi$ photoproduction cross sections in ultra-peripheral heavy-ion collisions at RHIC and at the LHC obtained in the framework of leading twist approximation of nuclear gluon shadowing. We showed the energy dependence of the coherent $\gamma {\rm Pb}\rightarrow J/\psi  {\rm Pb}$ cross section in the range of energies available for STAR and ALICE measurements and emphasized that the study of the coherent $J/\psi$ photoproduction in the range of rapidities of the STAR detector can be useful for understanding the tension between ALICE measurements and our predictions in the LTA approach. We also considered different classes of events with and without neutron emission and took into account the probability of neutron evaporation from the excited target nucleus  in the case of incoherent $J/\psi$ photoproduction. We showed that this effect significantly increases the cross section in the XnXn channel compared to the 0n0n channel. We also emphasized that the 0nXn selection in the incoherent $J/\psi$ photoproduction measurements allows one to tag target nuclei and disentangle low and high-energy contributions in UPC measurements.

\bibliographystyle{utphys}
\bibliography{main}

\providecommand{\href}[2]{#2}\begingroup\raggedright\begin{thebibliography}{10}

\bibitem{Baltz:2007kq}
A.~J. Baltz, ``{The Physics of Ultraperipheral Collisions at the LHC},''
  \href{http://dx.doi.org/10.1016/j.physrep.2007.12.001}{{\em Phys. Rept.}
  {\bfseries 458} (2008) 1--171},
  \href{http://arxiv.org/abs/0706.3356}{{\ttfamily arXiv:0706.3356 [nucl-ex]}}.

\bibitem{Contreras:2015dqa}
J.~G. Contreras and J.~D. Tapia~Takaki, ``{Ultra-peripheral heavy-ion
  collisions at the LHC},''
  \href{http://dx.doi.org/10.1142/S0217751X15420129}{{\em Int. J. Mod. Phys. A}
  {\bfseries 30} (2015) 1542012}.

\bibitem{STAR:2021wwq}
{ STAR} Collaboration, M.~Abdallah {\em et~al.}, ``{Probing the Gluonic
  Structure of the Deuteron with $J/\psi$ Photoproduction in d+Au
  Ultraperipheral Collisions},''
  \href{http://dx.doi.org/10.1103/PhysRevLett.128.122303}{{\em Phys. Rev.
  Lett.} {\bfseries 128} no.~12, (2022) 122303},
  \href{http://arxiv.org/abs/2109.07625}{{\ttfamily arXiv:2109.07625
  [nucl-ex]}}.

\bibitem{Klein:2016yzr}
S.~R. Klein, J.~Nystrand, J.~Seger, Y.~Gorbunov, and J.~Butterworth,
  ``{STARlight: A Monte Carlo simulation program for ultra-peripheral
  collisions of relativistic ions},''
  \href{http://dx.doi.org/10.1016/j.cpc.2016.10.016}{{\em Comput. Phys.
  Commun.} {\bfseries 212} (2017) 258--268},
  \href{http://arxiv.org/abs/1607.03838}{{\ttfamily arXiv:1607.03838
  [hep-ph]}}.

\bibitem{Frankfurt:2011cs}
L.~Frankfurt, V.~Guzey, and M.~Strikman, ``{Leading Twist Nuclear Shadowing
  Phenomena in Hard Processes with Nuclei},''
  \href{http://dx.doi.org/10.1016/j.physrep.2011.12.002}{{\em Phys. Rept.}
  {\bfseries 512} (2012) 255--393},
  \href{http://arxiv.org/abs/1106.2091}{{\ttfamily arXiv:1106.2091 [hep-ph]}}.

\bibitem{Cisek:2012yt}
A.~Cisek, W.~Schafer, and A.~Szczurek, ``{Exclusive coherent production of
  heavy vector mesons in nucleus-nucleus collisions at LHC},''
  \href{http://dx.doi.org/10.1103/PhysRevC.86.014905}{{\em Phys. Rev. C}
  {\bfseries 86} (2012) 014905},
  \href{http://arxiv.org/abs/1204.5381}{{\ttfamily arXiv:1204.5381 [hep-ph]}}.

\bibitem{Goncalves:2014wna}
V.~P. Goncalves, B.~D. Moreira, and F.~S. Navarra, ``{Investigation of
  diffractive photoproduction of $J/\Psi$ in hadronic collisions},''
  \href{http://dx.doi.org/10.1103/PhysRevC.90.015203}{{\em Phys. Rev. C}
  {\bfseries 90} no.~1, (2014) 015203},
  \href{http://arxiv.org/abs/1405.6977}{{\ttfamily arXiv:1405.6977 [hep-ph]}}.

\bibitem{Lappi:2013am}
T.~Lappi and H.~Mantysaari, ``{$J/|psi$ production in ultraperipheral Pb+Pb and
  $p$+Pb collisions at energies available at the CERN Large Hadron Collider},''
  \href{http://dx.doi.org/10.1103/PhysRevC.87.032201}{{\em Phys. Rev. C}
  {\bfseries 87} no.~3, (2013) 032201},
  \href{http://arxiv.org/abs/1301.4095}{{\ttfamily arXiv:1301.4095 [hep-ph]}}.

\bibitem{Cepila:2017nef}
J.~Cepila, J.~G. Contreras, and M.~Krelina, ``{Coherent and incoherent
  $\mathrm{J/}\psi$ photonuclear production in an energy-dependent hot-spot
  model},'' \href{http://dx.doi.org/10.1103/PhysRevC.97.024901}{{\em Phys. Rev.
  C} {\bfseries 97} no.~2, (2018) 024901},
  \href{http://arxiv.org/abs/1711.01855}{{\ttfamily arXiv:1711.01855
  [hep-ph]}}.

\bibitem{Luszczak:2019vdc}
A.~\L{}uszczak and W.~Sch\"afer, ``{Coherent photoproduction of $J/\psi$ in
  nucleus-nucleus collisions in the color dipole approach},''
  \href{http://dx.doi.org/10.1103/PhysRevC.99.044905}{{\em Phys. Rev. C}
  {\bfseries 99} no.~4, (2019) 044905},
  \href{http://arxiv.org/abs/1901.07989}{{\ttfamily arXiv:1901.07989
  [hep-ph]}}.

\bibitem{Bendova:2020hbb}
D.~Bendova, J.~Cepila, J.~G. Contreras, and M.~Matas, ``{Photonuclear $J/\psi$
  production at the LHC: Proton-based versus nuclear dipole scattering
  amplitudes},'' \href{http://dx.doi.org/10.1016/j.physletb.2021.136306}{{\em
  Phys. Lett. B} {\bfseries 817} (2021) 136306},
  \href{http://arxiv.org/abs/2006.12980}{{\ttfamily arXiv:2006.12980
  [hep-ph]}}.

\bibitem{Henkels:2020kju}
C.~Henkels, E.~G. de~Oliveira, R.~Pasechnik, and H.~Trebien, ``{Exclusive
  photoproduction of excited quarkonia in ultraperipheral collisions},''
  \href{http://dx.doi.org/10.1103/PhysRevD.102.014024}{{\em Phys. Rev. D}
  {\bfseries 102} no.~1, (2020) 014024},
  \href{http://arxiv.org/abs/2004.00607}{{\ttfamily arXiv:2004.00607
  [hep-ph]}}.

\bibitem{Frankfurt:2003wv}
L.~Frankfurt, M.~Strikman, and M.~Zhalov, ``{Coherent photoproduction from
  nuclei},'' {\em Acta Phys. Polon. B} {\bfseries 34} (2003) 3215--3254,
  \href{http://arxiv.org/abs/hep-ph/0304301}{{\ttfamily arXiv:hep-ph/0304301}}.

\bibitem{Cepila:2018zky}
J.~Cepila, J.~G. Contreras, M.~Krelina, and J.~D. Tapia~Takaki, ``{Mass
  dependence of vector meson photoproduction off protons and nuclei within the
  energy-dependent hot-spot model},''
  \href{http://dx.doi.org/10.1016/j.nuclphysb.2018.07.010}{{\em Nucl. Phys. B}
  {\bfseries 934} (2018) 330--340},
  \href{http://arxiv.org/abs/1804.05508}{{\ttfamily arXiv:1804.05508
  [hep-ph]}}.

\bibitem{Luszczak:2017dwf}
A.~\L{}uszczak and W.~Sch\"afer, ``{Incoherent diffractive photoproduction of
  $J/\psi$ and $\Upsilon$ on heavy nuclei in the color dipole approach},''
  \href{http://dx.doi.org/10.1103/PhysRevC.97.024903}{{\em Phys. Rev. C}
  {\bfseries 97} no.~2, (2018) 024903},
  \href{http://arxiv.org/abs/1712.04502}{{\ttfamily arXiv:1712.04502
  [hep-ph]}}.

\bibitem{Goncalves:2020vdp}
V.~P. Gon\c{c}alves, D.~E. Martins, and C.~R. Sena, ``{Coherent and incoherent
  $J/\Psi $ photoproduction in $Pb - Pb$ collisions at the LHC, HE-LHC and
  FCC},'' \href{http://dx.doi.org/10.1140/epja/s10050-021-00404-z}{{\em Eur.
  Phys. J. A} {\bfseries 57} no.~3, (2021) 82},
  \href{http://arxiv.org/abs/2007.13625}{{\ttfamily arXiv:2007.13625
  [hep-ph]}}.

\bibitem{Mantysaari:2017dwh}
H.~M\"antysaari and B.~Schenke, ``{Probing subnucleon scale fluctuations in
  ultraperipheral heavy ion collisions},''
  \href{http://dx.doi.org/10.1016/j.physletb.2017.07.063}{{\em Phys. Lett. B}
  {\bfseries 772} (2017) 832--838},
  \href{http://arxiv.org/abs/1703.09256}{{\ttfamily arXiv:1703.09256
  [hep-ph]}}.

\bibitem{Guzey:2018tlk}
V.~Guzey, M.~Strikman, and M.~Zhalov, ``{Nucleon dissociation and incoherent
  $J/\psi$ photoproduction on nuclei in ion ultraperipheral collisions at the
  Large Hadron Collider},''
  \href{http://dx.doi.org/10.1103/PhysRevC.99.015201}{{\em Phys. Rev. C}
  {\bfseries 99} no.~1, (2019) 015201},
  \href{http://arxiv.org/abs/1808.00740}{{\ttfamily arXiv:1808.00740
  [hep-ph]}}.

\bibitem{Good:1960ba}
M.~L. Good and W.~D. Walker, ``{Diffraction dissociation of beam particles},''
  \href{http://dx.doi.org/10.1103/PhysRev.120.1857}{{\em Phys. Rev.} {\bfseries
  120} (1960) 1857--1860}.

\bibitem{Frankfurt:2008vi}
L.~Frankfurt, M.~Strikman, D.~Treleani, and Weiss, ``{Evidence for color
  fluctuations in the nucleon in high-energy scattering},''
  \href{http://dx.doi.org/10.1103/PhysRevLett.101.202003}{{\em Phys. Rev.
  Lett.} {\bfseries 101} (2008) 202003},
  \href{http://arxiv.org/abs/0808.0182}{{\ttfamily arXiv:0808.0182 [hep-ph]}}.

\bibitem{Cepila:2016uku}
J.~Cepila, J.~G. Contreras, and J.~D. Tapia~Takaki, ``{Energy dependence of
  dissociative $\mathrm{J/}\psi$ photoproduction as a signature of gluon
  saturation at the LHC},''
  \href{http://dx.doi.org/10.1016/j.physletb.2016.12.063}{{\em Phys. Lett. B}
  {\bfseries 766} (2017) 186--191},
  \href{http://arxiv.org/abs/1608.07559}{{\ttfamily arXiv:1608.07559
  [hep-ph]}}.

\bibitem{ALICE:2012yye}
{ ALICE} Collaboration, B.~Abelev {\em et~al.}, ``{Coherent $J/\psi$
  photoproduction in ultra-peripheral Pb-Pb collisions at $\sqrt{s_{NN}} =
  2.76$ TeV},'' \href{http://dx.doi.org/10.1016/j.physletb.2012.11.059}{{\em
  Phys. Lett. B} {\bfseries 718} (2013) 1273--1283},
  \href{http://arxiv.org/abs/1209.3715}{{\ttfamily arXiv:1209.3715 [nucl-ex]}}.

\bibitem{ALICE:2013wjo}
{ ALICE} Collaboration, E.~Abbas {\em et~al.}, ``{Charmonium and $e^+e^-$ pair
  photoproduction at mid-rapidity in ultra-peripheral Pb-Pb collisions at
  $\sqrt{s_{\rm NN}}$=2.76 TeV},''
  \href{http://dx.doi.org/10.1140/epjc/s10052-013-2617-1}{{\em Eur. Phys. J. C}
  {\bfseries 73} no.~11, (2013) 2617},
  \href{http://arxiv.org/abs/1305.1467}{{\ttfamily arXiv:1305.1467 [nucl-ex]}}.

\bibitem{ALICE:2019tqa}
{ ALICE} Collaboration, S.~Acharya {\em et~al.}, ``{Coherent J/$\psi$
  photoproduction at forward rapidity in ultra-peripheral Pb-Pb collisions at
  $\sqrt{s_{\rm{NN}}}=5.02$ TeV},''
  \href{http://dx.doi.org/10.1016/j.physletb.2019.134926}{{\em Phys. Lett. B}
  {\bfseries 798} (2019) 134926},
  \href{http://arxiv.org/abs/1904.06272}{{\ttfamily arXiv:1904.06272
  [nucl-ex]}}.

\bibitem{ALICE:2021gpt}
{ ALICE} Collaboration, S.~Acharya {\em et~al.}, ``{Coherent $J/\psi$ and
  $\psi'$ photoproduction at midrapidity in ultra-peripheral Pb-Pb collisions
  at $\sqrt{s_{\mathrm{NN}}}~=~5.02$ TeV},''
  \href{http://dx.doi.org/10.1140/epjc/s10052-021-09437-6}{{\em Eur. Phys. J.
  C} {\bfseries 81} no.~8, (2021) 712},
  \href{http://arxiv.org/abs/2101.04577}{{\ttfamily arXiv:2101.04577
  [nucl-ex]}}.

\bibitem{ALICE:2021tyx}
{ ALICE} Collaboration, S.~Acharya {\em et~al.}, ``{First measurement of the
  |$t$|-dependence of coherent $J/\psi$ photonuclear production},''
  \href{http://dx.doi.org/10.1016/j.physletb.2021.136280}{{\em Phys. Lett. B}
  {\bfseries 817} (2021) 136280},
  \href{http://arxiv.org/abs/2101.04623}{{\ttfamily arXiv:2101.04623
  [nucl-ex]}}.

\bibitem{LHCb:2022ahs}
{ LHCb} Collaboration, R.~Aaij {\em et~al.}, ``{Study of exclusive
  photoproduction of charmonium in ultra-peripheral lead-lead collisions},''
  \href{http://dx.doi.org/10.1007/JHEP06(2023)146}{{\em JHEP} {\bfseries 06}
  (2023) 146}, \href{http://arxiv.org/abs/2206.08221}{{\ttfamily
  arXiv:2206.08221 [hep-ex]}}.

\bibitem{LHCb:2021bfl}
{ LHCb} Collaboration, R.~Aaij {\em et~al.}, ``{Study of coherent $J/\psi$
  production in lead-lead collisions at $ \sqrt{{\mathrm{s}}_{\mathrm{NN}}} $ =
  5 TeV},'' \href{http://dx.doi.org/10.1007/JHEP07(2022)117}{{\em JHEP}
  {\bfseries 07} (2022) 117}, \href{http://arxiv.org/abs/2107.03223}{{\ttfamily
  arXiv:2107.03223 [hep-ex]}}.

\bibitem{CMS:2016itn}
{ CMS} Collaboration, V.~Khachatryan {\em et~al.}, ``{Coherent $J/\psi$
  photoproduction in ultra-peripheral PbPb collisions at $\sqrt {s_{NN}} =$
  2.76 TeV with the CMS experiment},''
  \href{http://dx.doi.org/10.1016/j.physletb.2017.07.001}{{\em Phys. Lett. B}
  {\bfseries 772} (2017) 489--511},
  \href{http://arxiv.org/abs/1605.06966}{{\ttfamily arXiv:1605.06966
  [nucl-ex]}}.

\bibitem{Rebyakova:2011vf}
V.~Rebyakova, M.~Strikman, and M.~Zhalov, ``{Coherent $\rho$ and $J/\psi$
  photoproduction in ultraperipheral processes with electromagnetic
  dissociation of heavy ions at RHIC and LHC},''
  \href{http://dx.doi.org/10.1016/j.physletb.2012.03.041}{{\em Phys. Lett. B}
  {\bfseries 710} (2012) 647--653},
  \href{http://arxiv.org/abs/1109.0737}{{\ttfamily arXiv:1109.0737 [hep-ph]}}.

\bibitem{Guzey:2016piu}
V.~Guzey, E.~Kryshen, and M.~Zhalov, ``{Coherent photoproduction of vector
  mesons in ultraperipheral heavy ion collisions: Update for run 2 at the CERN
  Large Hadron Collider},''
  \href{http://dx.doi.org/10.1103/PhysRevC.93.055206}{{\em Phys. Rev. C}
  {\bfseries 93} no.~5, (2016) 055206},
  \href{http://arxiv.org/abs/1602.01456}{{\ttfamily arXiv:1602.01456
  [nucl-th]}}.

\bibitem{Guzey:2013jaa}
V.~Guzey, M.~Strikman, and M.~Zhalov, ``{Disentangling coherent and incoherent
  quasielastic $J/\psi$ photoproduction on nuclei by neutron tagging in
  ultraperipheral ion collisions at the LHC},''
  \href{http://dx.doi.org/10.1140/epjc/s10052-014-2942-z}{{\em Eur. Phys. J. C}
  {\bfseries 74} no.~7, (2014) 2942},
  \href{http://arxiv.org/abs/1312.6486}{{\ttfamily arXiv:1312.6486 [hep-ph]}}.

\bibitem{Baltz:2002pp}
A.~J. Baltz, S.~R. Klein, and J.~Nystrand, ``{Coherent vector meson
  photoproduction with nuclear breakup in relativistic heavy ion collisions},''
  \href{http://dx.doi.org/10.1103/PhysRevLett.89.012301}{{\em Phys. Rev. Lett.}
  {\bfseries 89} (2002) 012301},
  \href{http://arxiv.org/abs/nucl-th/0205031}{{\ttfamily
  arXiv:nucl-th/0205031}}.

\bibitem{Contreras:2016pkc}
J.~G. Contreras, ``{Gluon shadowing at small $x$ from coherent
  $\mathrm{J/}\psi$ photoproduction data at energies available at the CERN
  Large Hadron Collider},''
  \href{http://dx.doi.org/10.1103/PhysRevC.96.015203}{{\em Phys. Rev. C}
  {\bfseries 96} no.~1, (2017) 015203},
  \href{http://arxiv.org/abs/1610.03350}{{\ttfamily arXiv:1610.03350
  [nucl-ex]}}.

\bibitem{Pshenichnov:2011zz}
I.~A. Pshenichnov, ``{Electromagnetic excitation and fragmentation of
  ultrarelativistic nuclei},''
  \href{http://dx.doi.org/10.1134/S1063779611020067}{{\em Phys. Part. Nucl.}
  {\bfseries 42} (2011) 215--250}.

\bibitem{Broz:2019kpl}
M.~Broz, J.~G. Contreras, and J.~D. Tapia~Takaki, ``{A generator of forward
  neutrons for ultra-peripheral collisions: nOOn},''
  \href{http://dx.doi.org/10.1016/j.cpc.2020.107181}{{\em Comput. Phys.
  Commun.} {\bfseries 253} (2020) 107181},
  \href{http://arxiv.org/abs/1908.08263}{{\ttfamily arXiv:1908.08263
  [nucl-th]}}.

\bibitem{ALICE:2012aa}
{ ALICE} Collaboration, B.~Abelev {\em et~al.}, ``{Measurement of the Cross
  Section for Electromagnetic Dissociation with Neutron Emission in Pb-Pb
  Collisions at $\sqrt{s_{NN}}$ = 2.76 TeV},''
  \href{http://dx.doi.org/10.1103/PhysRevLett.109.252302}{{\em Phys. Rev.
  Lett.} {\bfseries 109} (2012) 252302},
  \href{http://arxiv.org/abs/1203.2436}{{\ttfamily arXiv:1203.2436 [nucl-ex]}}.

\bibitem{ALICE:2022iqi}
{ ALICE} Collaboration, S.~Acharya {\em et~al.}, ``{Neutron emission in
  ultraperipheral Pb-Pb collisions at $\sqrt {s_{NN}}$ = 5.02 TeV},''
  \href{http://dx.doi.org/10.1103/PhysRevC.107.064902}{{\em Phys. Rev. C}
  {\bfseries 107} no.~6, (2023) 064902},
  \href{http://arxiv.org/abs/2209.04250}{{\ttfamily arXiv:2209.04250
  [nucl-ex]}}.

\bibitem{ALICE:2021jnv}
{ ALICE} Collaboration, S.~Acharya {\em et~al.}, ``{First measurement of
  coherent $\rho^0$ photoproduction in ultra-peripheral Xe\textendash{}Xe
  collisions at $\sqrt{s_{\rm NN}} = 5.44$ TeV},''
  \href{http://dx.doi.org/10.1016/j.physletb.2021.136481}{{\em Phys. Lett. B}
  {\bfseries 820} (2021) 136481},
  \href{http://arxiv.org/abs/2101.02581}{{\ttfamily arXiv:2101.02581
  [nucl-ex]}}.

\bibitem{ALICE:2020ugp}
{ ALICE} Collaboration, S.~Acharya {\em et~al.}, ``{Coherent photoproduction of
  $\rho^{0}$ vector mesons in ultra-peripheral Pb-Pb collisions at $
  \sqrt{{\mathrm{s}}_{\mathrm{NN}}} $ = 5.02 TeV},''
  \href{http://dx.doi.org/10.1007/JHEP06(2020)035}{{\em JHEP} {\bfseries 06}
  (2020) 035}, \href{http://arxiv.org/abs/2002.10897}{{\ttfamily
  arXiv:2002.10897 [nucl-ex]}}.

\bibitem{ALICE:2015nbw}
{ ALICE} Collaboration, J.~Adam {\em et~al.}, ``{Coherent
  \ensuremath{\rho}$^{0}$ photoproduction in ultra-peripheral Pb-Pb collisions
  at $ \sqrt{s_{\mathrm{NN}}}=2.76 $ TeV},''
  \href{http://dx.doi.org/10.1007/JHEP09(2015)095}{{\em JHEP} {\bfseries 09}
  (2015) 095}, \href{http://arxiv.org/abs/1503.09177}{{\ttfamily
  arXiv:1503.09177 [nucl-ex]}}.

\bibitem{Strikman:2005ze}
M.~Strikman, M.~Tverskoy, and M.~Zhalov, ``{Neutron tagging of quasielastic
  $J/\psi$ photoproduction off nucleus in ultraperipheral heavy ion collisions
  at RHIC energies},''
  \href{http://dx.doi.org/10.1016/j.physletb.2005.08.083}{{\em Phys. Lett. B}
  {\bfseries 626} (2005) 72--79},
  \href{http://arxiv.org/abs/hep-ph/0505023}{{\ttfamily arXiv:hep-ph/0505023}}.

\bibitem{Guzey:2016qwo}
V.~Guzey, M.~Strikman, and M.~Zhalov, ``{Accessing transverse nucleon and gluon
  distributions in heavy nuclei using coherent vector meson photoproduction at
  high energies in ion ultraperipheral collisions},''
  \href{http://dx.doi.org/10.1103/PhysRevC.95.025204}{{\em Phys. Rev. C}
  {\bfseries 95} no.~2, (2017) 025204},
  \href{http://arxiv.org/abs/1611.05471}{{\ttfamily arXiv:1611.05471
  [hep-ph]}}.

\bibitem{Larionov:2018igy}
A.~B. Larionov and M.~Strikman, ``{Slow neutron production as a probe of hadron
  formation in high-energy $\gamma^*A$ reactions},''
  \href{http://dx.doi.org/10.1103/PhysRevC.101.014617}{{\em Phys. Rev. C}
  {\bfseries 101} no.~1, (2020) 014617},
  \href{http://arxiv.org/abs/1812.08231}{{\ttfamily arXiv:1812.08231
  [hep-ph]}}.

\bibitem{Ryskin:1992ui}
M.~G. Ryskin, ``{Diffractive $J/\psi$ electroproduction in LLA QCD},''
  \href{http://dx.doi.org/10.1007/BF01555742}{{\em Z. Phys. C} {\bfseries 57}
  (1993) 89--92}.

\bibitem{Jones:2013pga}
S.~P. Jones, A.~D. Martin, M.~G. Ryskin, and T.~Teubner, ``{Probes of the small
  $x$ gluon via exclusive $J/\psi$ and $\Upsilon$ production at HERA and the
  LHC},'' \href{http://dx.doi.org/10.1007/JHEP11(2013)085}{{\em JHEP}
  {\bfseries 11} (2013) 085}, \href{http://arxiv.org/abs/1307.7099}{{\ttfamily
  arXiv:1307.7099 [hep-ph]}}.

\bibitem{Guzey:2013qza}
V.~Guzey and M.~Zhalov, ``{Exclusive $J/{\psi}$ production in ultraperipheral
  collisions at the LHC: constrains on the gluon distributions in the proton
  and nuclei},'' \href{http://dx.doi.org/10.1007/JHEP10(2013)207}{{\em JHEP}
  {\bfseries 10} (2013) 207}, \href{http://arxiv.org/abs/1307.4526}{{\ttfamily
  arXiv:1307.4526 [hep-ph]}}.

\bibitem{H1:2013okq}
{ H1} Collaboration, C.~Alexa {\em et~al.}, ``{Elastic and Proton-Dissociative
  Photoproduction of J/psi Mesons at HERA},''
  \href{http://dx.doi.org/10.1140/epjc/s10052-013-2466-y}{{\em Eur. Phys. J. C}
  {\bfseries 73} no.~6, (2013) 2466},
  \href{http://arxiv.org/abs/1304.5162}{{\ttfamily arXiv:1304.5162 [hep-ex]}}.

\bibitem{Guzey:2013xba}
V.~Guzey, E.~Kryshen, M.~Strikman, and M.~Zhalov, ``{Evidence for nuclear gluon
  shadowing from the ALICE measurements of PbPb ultraperipheral exclusive
  $J/{\psi}$ production},''
  \href{http://dx.doi.org/10.1016/j.physletb.2013.08.043}{{\em Phys. Lett. B}
  {\bfseries 726} (2013) 290--295},
  \href{http://arxiv.org/abs/1305.1724}{{\ttfamily arXiv:1305.1724 [hep-ph]}}.

\bibitem{Camerini:1975cy}
U.~Camerini, J.~G. Learned, R.~Prepost, C.~M. Spencer, D.~E. Wiser, W.~Ash,
  R.~L. Anderson, D.~Ritson, D.~Sherden, and C.~K. Sinclair, ``{Photoproduction
  of the psi Particles},''
  \href{http://dx.doi.org/10.1103/PhysRevLett.35.483}{{\em Phys. Rev. Lett.}
  {\bfseries 35} (1975) 483}.

\bibitem{Binkley:1981kv}
M.~E. Binkley {\em et~al.}, ``{$J/\psi$ Photoproduction from 60 GeV/$c$ to 300
  GeV/$c$},'' \href{http://dx.doi.org/10.1103/PhysRevLett.48.73}{{\em Phys.
  Rev. Lett.} {\bfseries 48} (1982) 73}.

\bibitem{EuropeanMuon:1979nky}
{ European Muon} Collaboration, J.~J. Aubert {\em et~al.}, ``{Measurement of
  $J/\psi$ production in 280-GeV/c $\mu^+$ iron interactions},''
  \href{http://dx.doi.org/10.1016/0370-2693(80)90027-1}{{\em Phys. Lett. B}
  {\bfseries 89} (1980) 267--270}.

\bibitem{ZEUS:2002wfj}
{ ZEUS} Collaboration, S.~Chekanov {\em et~al.}, ``{Exclusive photoproduction
  of $J/\psi$ mesons at HERA},''
  \href{http://dx.doi.org/10.1007/s10052-002-0953-7}{{\em Eur. Phys. J. C}
  {\bfseries 24} (2002) 345--360},
  \href{http://arxiv.org/abs/hep-ex/0201043}{{\ttfamily arXiv:hep-ex/0201043}}.

\bibitem{H1:2005dtp}
{ H1} Collaboration, A.~Aktas {\em et~al.}, ``{Elastic $J/\psi$ production at
  HERA},'' \href{http://dx.doi.org/10.1140/epjc/s2006-02519-5}{{\em Eur. Phys.
  J. C} {\bfseries 46} (2006) 585--603},
  \href{http://arxiv.org/abs/hep-ex/0510016}{{\ttfamily arXiv:hep-ex/0510016}}.

\bibitem{ALICE:2014eof}
{ ALICE} Collaboration, B.~B. Abelev {\em et~al.}, ``{Exclusive
  $\mathrm{J/}\psi$ photoproduction off protons in ultra-peripheral p-Pb
  collisions at $\sqrt{s_{\rm NN}}=5.02$ TeV},''
  \href{http://dx.doi.org/10.1103/PhysRevLett.113.232504}{{\em Phys. Rev.
  Lett.} {\bfseries 113} no.~23, (2014) 232504},
  \href{http://arxiv.org/abs/1406.7819}{{\ttfamily arXiv:1406.7819 [nucl-ex]}}.

\bibitem{ALICE:2018oyo}
{ ALICE} Collaboration, S.~Acharya {\em et~al.}, ``{Energy dependence of
  exclusive $J/\psi $ photoproduction off protons in ultra-peripheral
  p\textendash{}Pb collisions at $\sqrt{s_{\mathrm {\scriptscriptstyle NN}}} =
  5.02$ TeV},'' \href{http://dx.doi.org/10.1140/epjc/s10052-019-6816-2}{{\em
  Eur. Phys. J. C} {\bfseries 79} no.~5, (2019) 402},
  \href{http://arxiv.org/abs/1809.03235}{{\ttfamily arXiv:1809.03235
  [nucl-ex]}}.

\bibitem{LHCb:2014acg}
{ LHCb} Collaboration, R.~Aaij {\em et~al.}, ``{Updated measurements of
  exclusive $J/\psi$ and $\psi$(2S) production cross-sections in pp collisions
  at $\sqrt{s}=7$ TeV},''
  \href{http://dx.doi.org/10.1088/0954-3899/41/5/055002}{{\em J. Phys. G}
  {\bfseries 41} (2014) 055002},
  \href{http://arxiv.org/abs/1401.3288}{{\ttfamily arXiv:1401.3288 [hep-ex]}}.

\bibitem{LHCb:2018rcm}
{ LHCb} Collaboration, R.~Aaij {\em et~al.}, ``{Central exclusive production of
  $J/\psi$ and $\psi(2S)$ mesons in $pp$ collisions at $\sqrt{s}=13~$TeV},''
  \href{http://dx.doi.org/10.1007/JHEP10(2018)167}{{\em JHEP} {\bfseries 10}
  (2018) 167}, \href{http://arxiv.org/abs/1806.04079}{{\ttfamily
  arXiv:1806.04079 [hep-ex]}}.

\bibitem{Glauber:1959}
R.~J. Glauber, ``High-energy collision theory,'' in {\em Lectures in
  Theoretical Physics}, pp.~315--414.
\newblock Interscience, New York, 1959.

\bibitem{Gribov:1968jf}
V.~N. Gribov, ``{Glauber corrections and the interaction between high-energy
  hadrons and nuclei},'' {\em Sov. Phys. JETP} {\bfseries 29} (1969) 483--487.

\bibitem{Gribov:2009zz}
V.~N. Gribov, {\em {Strong interactions of hadrons at high emnergies: Gribov
  lectures on Theoretical Physics}}.
\newblock Cambridge University Press, 10, 2012.

\bibitem{H1:2002yab}
{ H1} Collaboration, C.~Adloff {\em et~al.}, ``{Diffractive photoproduction of
  psi(2S) mesons at HERA},''
  \href{http://dx.doi.org/10.1016/S0370-2693(02)02275-X}{{\em Phys. Lett. B}
  {\bfseries 541} (2002) 251--264},
  \href{http://arxiv.org/abs/hep-ex/0205107}{{\ttfamily arXiv:hep-ex/0205107}}.

\bibitem{ZEUS:2022sxn}
{ ZEUS} Collaboration, I.~Abt {\em et~al.}, ``{Measurement of the cross-section
  ratio \ensuremath{\sigma}$_{\psi(2S)}$/\ensuremath{\sigma}$_{J/\psi(1S)}$ in
  exclusive photoproduction at HERA},''
  \href{http://dx.doi.org/10.1007/JHEP12(2022)164}{{\em JHEP} {\bfseries 12}
  (2022) 164}, \href{http://arxiv.org/abs/2206.13343}{{\ttfamily
  arXiv:2206.13343 [hep-ex]}}.

\bibitem{Guzey:2022qvc}
V.~Guzey, E.~Kryshen, M.~Strikman, and M.~Zhalov, ``{Photoproduction of
  $J/\psi$ in d+Au ultraperipheral collisions at the BNL Relativistic Heavy Ion
  Collider},'' \href{http://dx.doi.org/10.1103/PhysRevC.106.064909}{{\em Phys.
  Rev. C} {\bfseries 106} no.~6, (2022) 064909},
  \href{http://arxiv.org/abs/2206.12120}{{\ttfamily arXiv:2206.12120
  [hep-ph]}}.

\bibitem{E665:1995utr}
{ E665} Collaboration, M.~R. Adams {\em et~al.}, ``{Nuclear decay following
  deep inelastic scattering of 470-GeV muons},''
  \href{http://dx.doi.org/10.1103/PhysRevLett.74.5198}{{\em Phys. Rev. Lett.}
  {\bfseries 74} (1995) 5198--5201}. [Erratum: Phys.Rev.Lett. 80, 2020--2021
  (1998)].

\bibitem{Strikman:1996ak}
M.~I. Strikman, M.~G. Tverskoi, and M.~B. Zhalov, ``{Soft neutron productions:
  A Window to the final state interactions in DIS},'' in {\em {Workshop on
  Future Physics at HERA}}.
\newblock 5, 1996.
\newblock \href{http://arxiv.org/abs/nucl-th/9609055}{{\ttfamily
  arXiv:nucl-th/9609055}}.

\bibitem{Strikman:1998cc}
M.~Strikman, M.~G. Tverskoii, and M.~B. Zhalov, ``{Soft neutron production in
  DIS: A Window to the final state interactions},''
  \href{http://dx.doi.org/10.1016/S0370-2693(99)00627-9}{{\em Phys. Lett. B}
  {\bfseries 459} (1999) 37--42},
  \href{http://arxiv.org/abs/nucl-th/9806099}{{\ttfamily
  arXiv:nucl-th/9806099}}.

\end{thebibliography}\endgroup

\clearpage
\newpage

\clearpage
\newpage

\clearpage
\newpage

\clearpage
\newpage

\clearpage
\newpage

\clearpage
\newpage

\clearpage
\newpage

\clearpage
\newpage

\clearpage
\newpage

\end{document}